\documentclass[pdflatex,sn-mathphys-num]{sn-jnl}
\usepackage{graphicx}
\usepackage{bm}
\usepackage{IEEEtrantools}
\usepackage{amsthm}
\usepackage{algorithm}
\usepackage{algpseudocode}
\usepackage{siunitx}
\usepackage{caption}
\usepackage{subcaption}
\usepackage{tabularx}
\usepackage{float}
\usepackage{amsmath}
\usepackage{mathrsfs}
\usepackage{hyperref}
\usepackage{threeparttablex}

\newcolumntype{b}{X}
\newcolumntype{k}{>{\hsize=.3\hsize}X}
\newcolumntype{m}{>{\hsize=.25\hsize}X}
\newcolumntype{s}{>{\hsize=.5\hsize}X}
\newcolumntype{u}{>{\hsize=.45\hsize}X}
\newcolumntype{q}{>{\hsize=.55\hsize}X}

\begin{document}

\title[Scintillation pulse characterization]{Scintillation pulse characterization with spectrum-inspired temporal neural networks: case studies on particle detector signals}

\author*[1,2]{\fnm{Pengcheng} \sur{Ai}}\email{aipc@ccnu.edu.cn}

\author[1,2]{\fnm{Xiangming} \sur{Sun}}

\author[3]{\fnm{Zhi} \sur{Deng}}

\author[3]{\fnm{Xinchi} \sur{Ran}}

\affil*[1]{\orgdiv{PLAC, Key Laboratory of Quark and Lepton Physics (MOE)}, \orgname{Central China Normal University}, \orgaddress{\street{No. 152 Luoyu Road}, \city{Wuhan}, \postcode{430079}, \state{Hubei}, \country{China}}}

\affil[2]{\orgname{Hubei Provincial Engineering Research Center of Silicon Pixel Chip \& Detection Technology}, \orgaddress{\street{No. 152 Luoyu Road}, \city{Wuhan}, \postcode{430079}, \state{Hubei}, \country{China}}}

\affil[3]{\orgdiv{Key Laboratory of Particle and Radiation Imaging (MOE), Department of Engineering Physics}, \orgname{Tsinghua University}, \orgaddress{\street{No. 30 Shuangqing Road}, \city{Beijing}, \postcode{100084}, \country{China}}}

\abstract{Particle detectors based on scintillators are widely used in high-energy physics and astroparticle physics experiments, nuclear medicine imaging, industrial and environmental detection, etc. Precisely extracting scintillation signal characteristics at the event level is important for these applications, not only in respect of understanding the scintillator itself, but also kinds and physical property of incident particles. Recent researches demonstrate data-driven neural networks surpass traditional statistical methods, especially when the analytical form of signals is hard to obtain, or noise is significant. However, most densely connected or convolution-based networks fail to fully exploit the spectral and temporal structure of scintillation signals, leaving large space for performance improvement. In this paper, we propose a network architecture specially tailored for scintillation pulse characterization based on previous works on time series analysis. The core insight is that, by directly applying Fast Fourier Transform on original signals and utilizing different frequency components, the proposed network architecture can serve as a lightweight and enhanced representation learning backbone. We prove our idea in two case studies: (a) simulation data generated with the setting of the LUX dark matter detector, and (b) experimental electrical signals with fast electronics to emulate scintillation variations for the NICA/MPD calorimeter. The proposed model achieves significantly better results than the reference model in literature and densely connected models and demonstrates higher cost-efficiency than conventional machine learning methods.}

%
%
%
\maketitle
%
%

\section{Introduction}

Particle detectors based on scintillators are widely used in high-energy physics and astroparticle physics experiments \cite{Lee2024}, nuclear medicine imaging \cite{Sun2023,Osmanagaoglu2024}, industrial and environmental detection \cite{PANDYA2025170177,Saeidi2025}, etc. The scintillation light is converted to electrical signals by transducers like photomultipliers (PMT) or silicon photomultipliers and then processed by subsequent circuits. With the rapid development of electronic devices and integrated circuits, nowadays high bandwidth front-end electronics and high sampling rate analog-to-digital converters become affordable even for medium or large detector systems. This generates a considerable amount of raw data coming out of digitizers. The data explosion in nuclear instrumentation offers opportunities as well as challenges for the community, which gives rise to researches on software algorithms \cite{FU2018410,Ai_2019} and related online hardware implementation \cite{Aarrestad_2021,Khoda_2023,10005128} for signal processing.

Neural networks, experiencing a renaissance after deep learning, have been utilized as a major technique for intelligent processing of scintillation signals. Depending on requirements of applications, neural networks have been used as classifiers in n/$\gamma$ discrimination \cite{DOUCET2020161201,Zhao_2023,LEE2024169638}, pulse shape discrimination \cite{Dutta_2023,Jung_2023,Griffiths_2020,Cheng2024}, particle identification \cite{CARLINI2024169369,KIM201983}, anomaly detection \cite{Angloher2023}, etc. Furthermore, in scenarios of precision measurement, neural networks demonstrate improved performance in pile-up discrimination \cite{KIM2023110880,9667358,REGADIO2021165403}, time-of-flight estimation \cite{Berg_2018,Onishi_2022}, pulse timing \cite{10038575,Ai_2023}, pulse height estimation \cite{KIM2023110880,9667358,REGADIO2021165403} and energy extraction \cite{HESHMATI2022110265,TAJIK2024111375,Jiang2025}. The pervasive use of neural networks shows its feasibility as an efficient feature extractor from digitized scintillation signals.

Most of neural networks in the above literature are densely connected or convolution-based. Though these architectures are theoretically proved to be universal approximators, or experimentally validated in areas like computer vision, there are evidences showing that the potential of neural networks has not been fully unleashed for scintillation signals. For example, in pulse timing, Cram\'er Rao lower bound of timing performance is deduced for radiation detectors and neural networks are compared against it \cite{Ai_2021}. There is still some gap between the lower bound and the performance of neural networks when pulse variation is significant or sampling rate is low. This sheds light on explorations on more effective and universal network architectures for scintillation signals.

In this work, we demonstrate that the neural network architecture inspired by spectral analysis of scintillation signals can be a justified choice for multiple regression and classification tasks. This network architecture originates from \emph{TimesNet} \cite{DBLP:conf/iclr/WuHLZ0L23}, a recent advance in time series analysis, but some important adjustments are incorporated to make the model \emph{lightweight} and \emph{enhanced} in terms of performance. We call the new model \emph{TimesNet-LE}. Rather than making a horizontal comparison between different network architectures, this work aims at demonstrating how the performance of simple networks can be boosted with representation learning implemented by the \emph{TimesNet-LE} backbone. The major innovations and contributions of the paper include:

\begin{itemize}
	\item A highly compact and effective network architecture, \emph{TimesNet-LE}, for feature extraction from scintillation signals: the network architecture comprises blocks of temporal multi-period convolution based on dominant periods in the spectrum.
	\item Application of the above network architecture on particle detector signals, demonstrating that it can serve as an efficient representation learning backbone to support complex tasks from multimodal regression to multi-class classification.
	\item Simulation with the liquid xenon scintillator of the LUX dark matter detector, and the experiment with fast electronics of the NICA/MPD calorimeter, both of which show the performance boost of the method in a wide area.
\end{itemize}

\section{Method}

\subsection{Spectrum-inspired multi-period convolution}
\label{sec:multi-period-conv}

\begin{figure}[htb]
	\centering
	\includegraphics[width=0.98\textwidth]{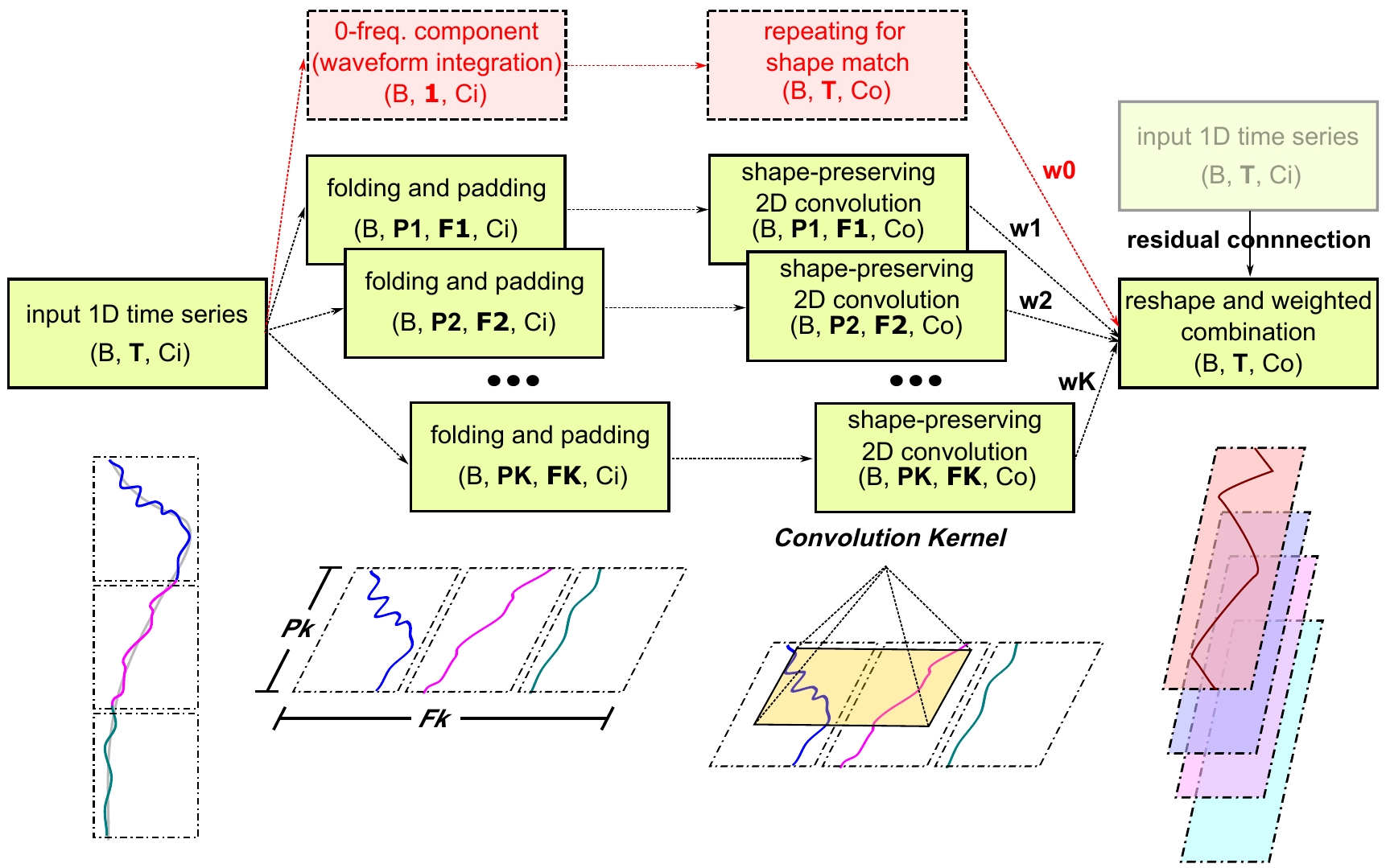}
	\caption{The functional diagram of multi-period convolution blocks. The symbol in the parentheses at the bottom of each square denotes the tensor shape in each step. The red paths and squares are components not in original \emph{TimesNet}.}
	\label{fig:timesnet-block}
\end{figure}

Scintillation signals can be viewed as a kind of time series with data recorded at fixed time intervals. From the perspective of frequency domain, its spectrum is periodic, and can be discretized with Fast Fourier Transform (FFT). By finding dominant frequencies (or corresponding periods) in the frequency domain and reshaping the time series with the frequencies (or periods) into two-dimensional (2D) maps, feature extraction on these 2D maps can be much more effective. This is the main idea behind \emph{TimesNet} \cite{DBLP:conf/iclr/WuHLZ0L23}, a state-of-the-art model for time series forecasting, imputation, classification, etc.

In Fig. \ref{fig:timesnet-block}, we show the inner structure of multi-period convolution blocks (called \emph{TimesBlock} in \cite{DBLP:conf/iclr/WuHLZ0L23}), the building block of \emph{TimesNet}. the input one-dimensional (1D) time series is folded and padded when necessary at several dominant periods ($(Pk, Fk),\ k=1,2,\dots,K$) at its length dimension ($T$). A group of convolution kernels with different sizes act on folded 2D maps and generate several output channels ($Co$) from input channels ($Ci$) while keeping the shape of 2D maps as the same. Finally, the 2D maps from different dimensions are reshaped again into the previous length ($T$) and combined with weight values of corresponding amplitudes in the FFT spectrum. Residual connection is applied between input and output (when $Ci \neq Co$, broadcasting may happen).

It should be noted that, in original \emph{TimesNet}, the 0-frequency component is directly abandoned, since it cannot be reshaped into a 2D map with zero length at its side. However, this component can be information-rich for scintillation signals. This path is added in red color in Fig. \ref{fig:timesnet-block} and will be explained in the subsequent section.

\subsection{TimesNet-LE}

\begin{figure}[htb]
	\centering
	\includegraphics[width=0.98\textwidth]{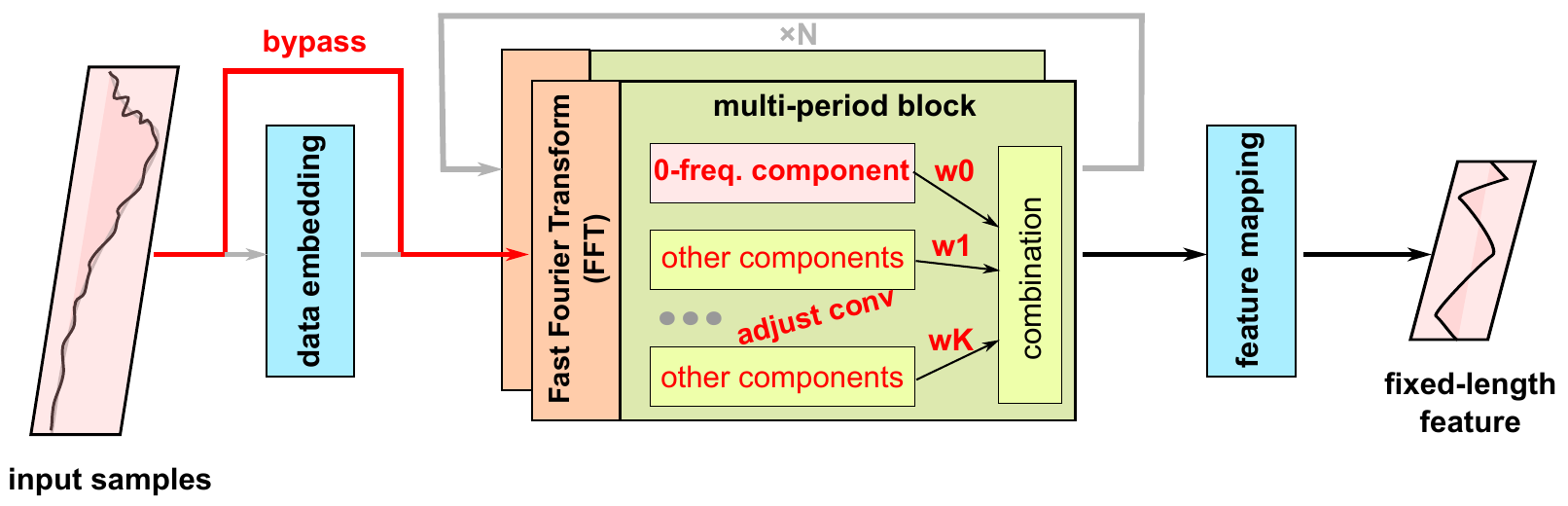}
	\caption{The functional diagram of \emph{TimesNet-LE}. The path and texts in red color are main differences between \emph{TimesNet} and \emph{TimesNet-LE}.}
	\label{fig:timesnet-le}
\end{figure}

Although \emph{TimesNet} shows promising performance on several time series benchmarks, the predefined universal model has its limitations when applied to scintillation signals. Scintillation pulses are time series with a single channel, and their dominant frequencies lie heavily in the low-frequency region of the spectrum. Therefore, we cannot regard them as common time series in nature and society, such as temperature or consumption index.

To fit in these particular conditions, we propose \emph{TimesNet-LE}, which is short for \emph{TimesNet Lightweight and Enhanced}, as the major network architecture discussed in the paper. The adjustments of \emph{TimesNet-LE}, shown in Fig. \ref{fig:timesnet-le}, include:

\paragraph{Directly applying FFT on original signals without data embedding.} The data embedding layer in original \emph{TimesNet} consists of a densely connected token embedding and a fixed position embedding. The dense connection brings about additional complexities. To streamline the model and make it more interpretable, the data embedding is bypassed in \emph{TimesNet-LE}. As a result, the computed FFT spectrum in the first multi-period block strictly corresponds to the spectrum of the input signal. This makes the following adjustments permissible.

\paragraph{Including the zero-frequency component.} The original \emph{TimesNet} abandons the 0-frequency component for consistency in the multi-period reshaping. However, it is vital to include it for utilizing the direct current of the signal. Therefore, we include the 0-frequency component as an additional path in the multi-period block (shown at the top of Fig. \ref{fig:timesnet-block}). We integrate the input signal at its length dimension, and repeat the integration to match the whole length ($T$) and multiple channels ($Co$). In this way, the 0-frequency component is effectively exploited.

\paragraph{Adjusting the convolution scheme for low-frequency components.} In the original \emph{TimesNet}, the convolution kernels for frequency components are relatively big, spanning a wide area. However, as aforementioned, the low-frequency components dominate the spectrum, so the big convolution kernels seem redundant. In view of this, we reduce the convolution kernels for low-frequency components for more effective convolution.

\paragraph{Unbiasedly re-weighting features from different frequencies.} The original \emph{TimesNet} use amplitudes in the FFT spectrum as weight values when combining different frequency components. Hence, it will emphasize the component with the large amplitude, and neglect the component with the small amplitude. This is reasonable when the FFT spectrum is intrinsically multimodal (with multiple maximums), which is not the case for scintillation signals. Besides, we tend to care more about the small-amplitude components when their information is not negligible. In \emph{TimesNet-LE}, we propose to use unbiased weight values for all frequency components for better taking advantage of intricate structures in components with relatively small amplitudes.

\subsection{Representation learning schemes}
\label{sec:rep-learn}

\begin{figure}[htb]
	\centering
	\includegraphics[width=0.7\textwidth]{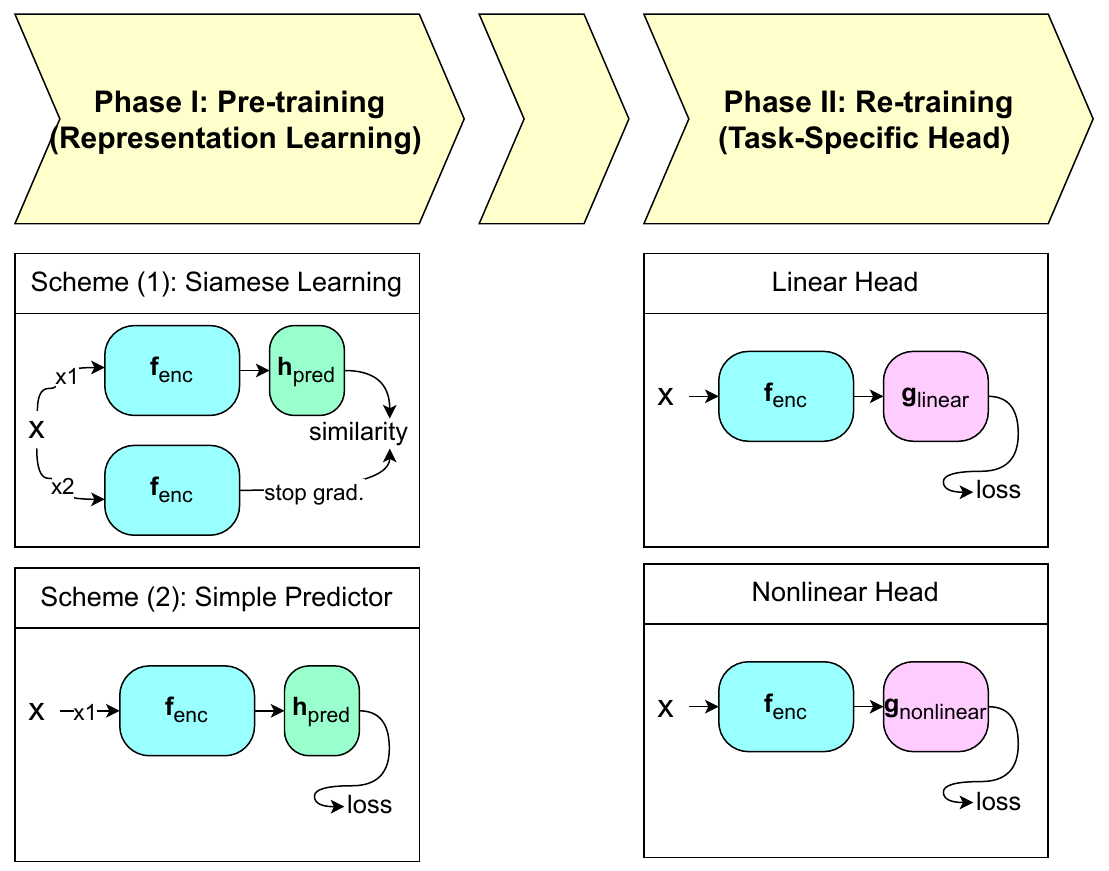}
	\caption{The functional diagram of representation learning schemes.}
	\label{fig:rep-learn}
\end{figure}

The performance of \emph{TimesNet} is validated on multiple tasks of time series analysis. It is intriguing to think of similar networks as backbones for representation learning. In this study, we have employed two representation learning schemes (shown in Fig. \ref{fig:rep-learn}):

\begin{itemize}
	\item[(1)] \emph{Siamese Learning}: We use SimSiam \cite{DBLP:conf/cvpr/ChenH21}, a self-supervised learning method, to pre-train the encoder model as the backbone. Two augmented versions of input data are fed into two encoder networks with weight sharing. One branch has a predictor network, and gradient back-propagation is stopped in another branch. The similarity loss function is computed between two branches for self-supervised training.
	\item[(2)] \emph{Encoder with Simple Predictor}: In this scheme, a simple predictor network follows the encoder. A single augmented version of input data is fed to the encoder. A commonly used loss function is applied at the output of the predictor network. The task used in pre-training can be the same as or different from the task in re-training.
\end{itemize}

When pre-training is done, re-training task-specific heads on feature embedding in the representation learning fits a particular purpose. The head networks are kept as simple as possible. In this study we experiment with two heads: a linear head with no hidden layer and a nonlinear head with only one hidden layer. In most cases, the weights in the encoder network are fixed during the re-training phase.

\section{Case studies}

\subsection{LUX dark matter detector}

\subsubsection{Simulation setup}

\begin{figure}[htb]
	\centering
	\begin{subfigure}[b]{0.48\textwidth}
		\centering
		\includegraphics[width=\textwidth]{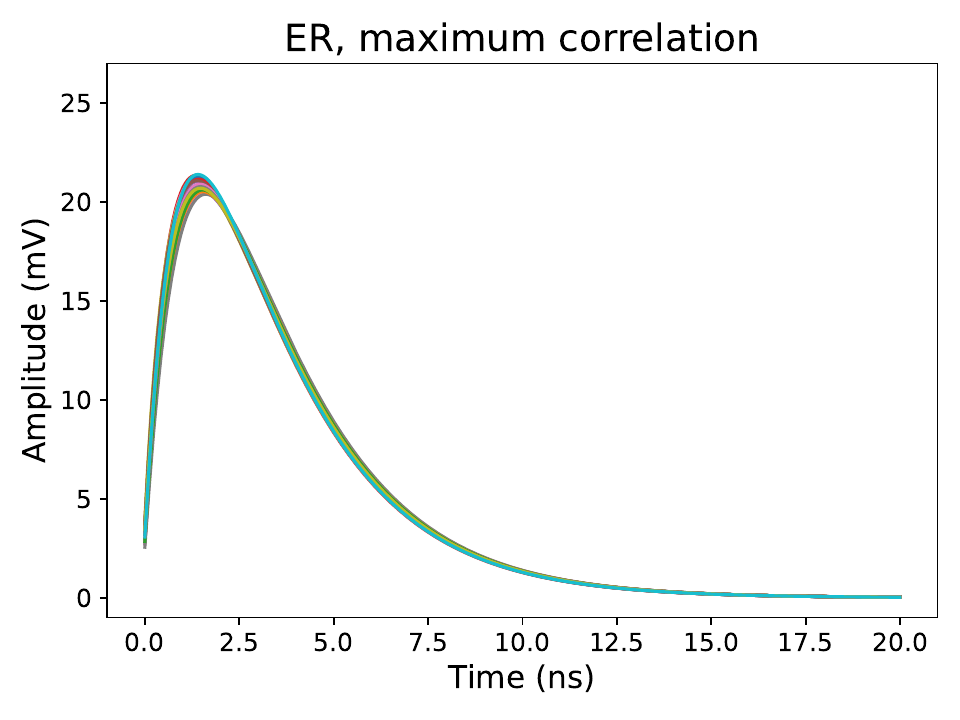}
		\caption{Backgrounds of electron recoils (ER)}
	\end{subfigure}
	\begin{subfigure}[b]{0.48\textwidth}
		\centering
		\includegraphics[width=\textwidth]{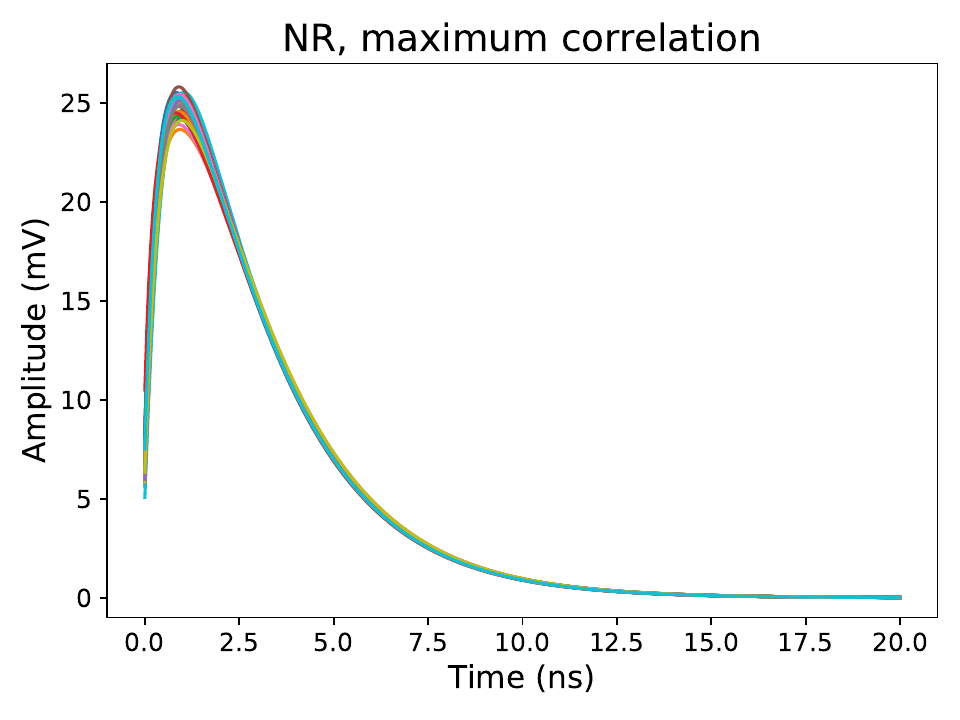}
		\caption{Signals of nuclear recoils (NR)}
	\end{subfigure}
	\caption{Simulation data of scintillation signals in the LUX experiment. Data with maximum correlation between parameters are visualized. We select the first 20 ns and plot 30 curves in each sub-figure.}
	\label{fig:data-viz-lux}
\end{figure}

The Large Underground Xenon (LUX) experiment \cite{AKERIB2013111,PhysRevD.97.112002} is a project searching for data matter. The instrument of the LUX experiment contains a huge cylindrical chamber of liquid xenon, with PMTs arranged as arrays at two sides of the chamber. Scintillation light is produced by the self-trapping of excited xenon atoms, or gamma/beta radiation interacting with xenon. The former are the \emph{signal} events of nuclear recoils (NR), while the latter are the major \emph{background} events of electron recoils (ER). Both signal and background events are detected by PMTs and processed by subsequent electronics. In Fig. \ref{fig:data-viz-lux}, we show simulated ER and NR events received by PMTs before low-pass filtering.

It can be seen in Fig. \ref{fig:data-viz-lux} that ER and NR events have distinct characteristics. Although the original intention in \cite{PhysRevD.97.112002} is to discriminate between ER and NR events, in this work, we aim to predict the event-related parameters of scintillation signals. The problem is intrinsically multimodal because ER and NR events span different intervals in the parameter range. Precision measurement of scintillation-related parameters is meaningful for better interpretation of experimental data, and serves as a probe to observe the property of weakly interacting massive particles, a major proposition by the dark matter theory.

The scintillation emission consists of a fast component and a slow component. In this simulation, two parameters of scintillation signals are researched: the time coefficient of the slow component, and the ratio of the slow component. The fast component is too challenging to predict because its time coefficient is on the same order of magnitude as the sampling period, and its ratio is too small. We use a low-pass filter on scintillation signals for pulse shaping, and vary the amplitude to simulate the variation of electronic gains. Besides, noise on the source side before low-pass filtering (\emph{correlated} noise) and on the terminal side after low-pass filtering (\emph{uncorrelated} noise) are separately considered. The rigorous mathematical form of the generated waveform and the whole simulation process are discussed in detail in Appendix \ref{sec:detail-sim}.

In should be noted that, under the above conditions, using traditional analyzing methods like curve fitting is inefficient because of the complicated effects of scintillation emission (equation \ref{equ:scint-emi}), optical process (equation \ref{equ:optic-proc}), electronics (equation \ref{equ:filter}) and noise (equation \ref{equ:noise}). Directly combining these factors will result in a frequency-domain system function with 6 poles and several zeros. The locations of poles and zeros will change with input parameters, which can be quite ill-conditioned when solving the reverse optimization problem. Besides, considering noise correlation will incur a large covariance matrix in curve fitting, which is very compute-intensive. On the other hand, neural networks become a natural choice if sufficient data are generated.

\subsubsection{Results}

In simulation, the 10-giga samples per second (GSPS) raw waveform is downsampled with a ratio 20:1, and a section of 256 samples is randomly selected on the downsampled waveform. This gives an input time series with the 2-ns period, or equivalently the 500-mega samples per second (MSPS) rate, and 256 elements. We have generated 10,000 ER events and 10,000 NR events, forming a dataset with 20,000 examples in total. Then the dataset is divided into a training set with 14,000 examples, a validation set with 2000 examples and a test set with 4000 examples.

We totally consider 7 network architectures grouped by 3 classes:

\begin{enumerate}
	\item Head directly on samples. In this class, a network with only a linear output layer (tagged \emph{linear head on samples}) or a network with a hidden layer and a linear output layer (tagged \emph{nonlinear head on samples}) is used to predict two parameters based on waveform samples as input. Training is performed in an end-to-end manner.
	\item Head on features predicted by \emph{TimesNet}. In this class, we first use representation learning scheme (1) (see section \ref{sec:rep-learn}) to pre-train \emph{TimesNet} as the backbone\footnote{We find that directly training a \emph{TimesNet} network in a task-specific, end-to-end manner within our simulation environment tends to generate results varied run by run, which indicates trapping in sub-optimal solutions and overfitting.}. Then the backbone is kept fixed, and a linear network (tagged \emph{linear head on 32 feat.} or \emph{linear head on 256 feat.} with different feature sizes) or a nonlinear network (tagged \emph{nonlinear head on 256 feat.}) is trained upon the backbone.
	\item Head on features predicted by \emph{TimesNet-LE}. In this class, we first use representation learning scheme (2) to pre-train \emph{TimesNet-LE} as the backbone. Then the backbone is kept fixed, and a linear network (tagged \emph{linear head on 32 feat.}) or a nonlinear network (tagged \emph{nonlinear head on 32 feat.}) is trained upon the backbone.
\end{enumerate}

\begin{table}[htb]
	\centering
	\caption{Numbers of trainable parameters in different network architectures in simulation.}
	\label{tab:param-sim}
	\small
	\begin{tabularx}{\textwidth}{bqu}
		\hline
		\textbf{network architecture} & \textbf{\#param (encoder)} & \textbf{\#param (head)} \\
		\hline
		linear head on samples & None & 0.8k \\
		nonlinear head on samples & None & 49.5k \\
		linear head on 32 feat. (\emph{TimesNet}) & 594.6k & $<$0.1k \\
		linear head on 256 feat. (\emph{TimesNet}) & 652.2k & 0.8k \\
		nonlinear head on 256 feat. (\emph{TimesNet}) & 652.2k & 49.5k \\
		linear head on 32 feat. (\emph{TimesNet-LE}) & 27.4k & $<$0.1k \\
		nonlinear head on 32 feat. (\emph{TimesNet-LE}) & 27.4k & 0.5k \\
		\hline
	\end{tabularx}
\end{table}

In table \ref{tab:param-sim}, we give the numbers of trainable parameters in these different network architectures in simulation. It can be seen that, for architectures with backbones, the major trainable parameters lie in the encoder; besides, in comparison with \emph{TimesNet}, \emph{TimesNet-LE} greatly reduces the number of trainable parameters. This is in line with the assumption that \emph{TimesNet-LE} is a lightweight model. More details about the configuration and network architectures can be found in Appendix \ref{sec:detail-net-arch}.

\begin{figure}[htb]
	\centering
	\begin{subfigure}[b]{0.48\textwidth}
		\centering
		\includegraphics[width=\textwidth]{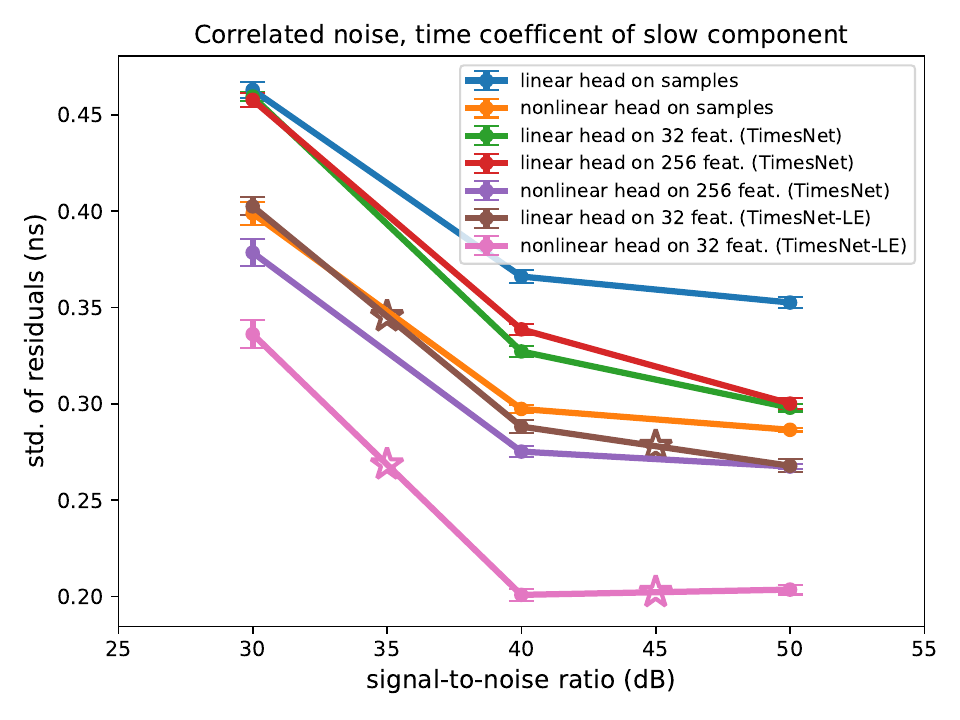}
		\caption{Time coefficient, correlated noise}
		\label{fig:lux-perf-tc-corr}
	\end{subfigure}
	\begin{subfigure}[b]{0.48\textwidth}
		\centering
		\includegraphics[width=\textwidth]{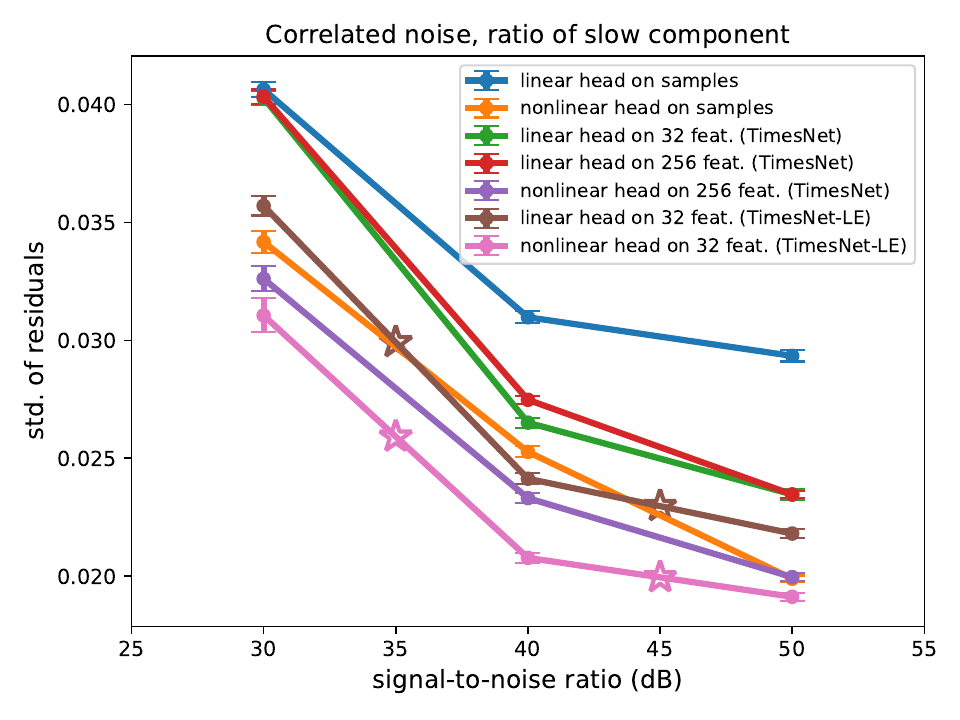}
		\caption{Ratio, correlated noise}
		\label{fig:lux-perf-ratio-corr}
	\end{subfigure}
	\begin{subfigure}[b]{0.48\textwidth}
		\centering
		\includegraphics[width=\textwidth]{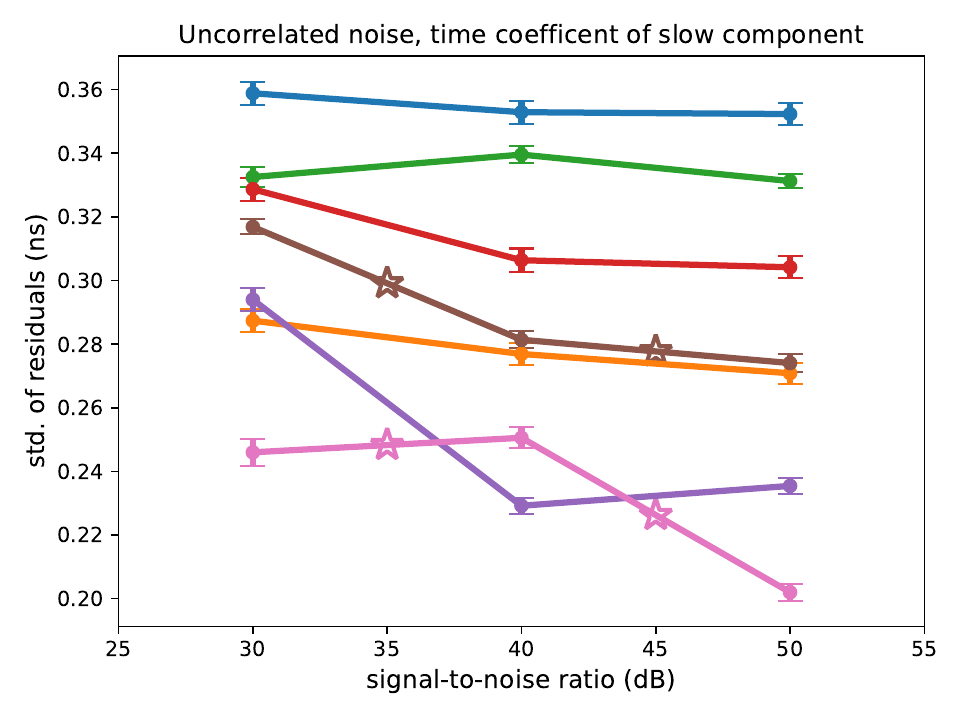}
		\caption{Time coefficient, uncorrelated noise}
		\label{fig:lux-perf-tc-uncorr}
	\end{subfigure}
	\begin{subfigure}[b]{0.48\textwidth}
		\centering
		\includegraphics[width=\textwidth]{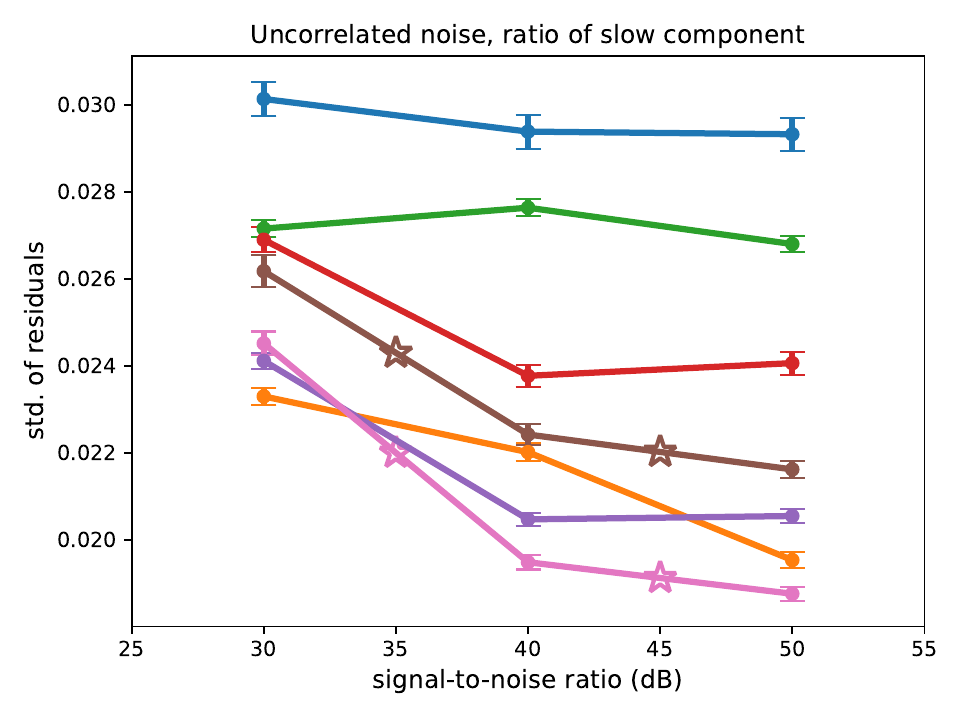}
		\caption{Ratio, uncorrelated noise}
		\label{fig:lux-perf-ratio-uncorr}
	\end{subfigure}
	\caption{Performance measures (standard deviation of residuals between the ground-truth and the predicted) of parameter estimation on simulation data with \emph{correlated} or \emph{uncorrelated} noise at different signal-to-noise ratios. Network architectures with \emph{TimesNet-LE} are marked with stars. Sub-figures in the bottom row share the same legend with those in the top row.}
	\label{fig:lux-perf}
\end{figure}

To evaluate different network architectures at different levels of signal quality, we generate time series with signal-to-noise ratio (SNR) at 30 dB, 40 dB and 50 dB, with the larger value indicating better signal quality. For each SNR, we independently train each network architecture with fixed epochs when the loss is substantially reduced.

The main results are shown in Fig. \ref{fig:lux-perf}. We separately discuss conditions with \emph{correlated} noise and \emph{uncorrelated} noise as follows:

\paragraph{Correlated noise.} The noise originates from the source side and becomes correlated after low-pass filtering. In this condition, all network architectures show steadily improved performance when signal quality gets better. By examining the curves in Fig. \ref{fig:lux-perf-tc-corr} and Fig. \ref{fig:lux-perf-ratio-corr}, we find that, for the same class of networks, the architecture with the nonlinear head performs better than the architecture with the linear head, which indicates the importance of nonlinear mapping in these regression tasks. We can also find that architectures with the backbone perform better than architectures without the backbone when they share the same heads, indicating the positive effects of \emph{TimesNet} or \emph{TimesNet-LE} in representation learning. Furthermore, it can be seen that \emph{TimesNet-LE} gives the largest performance boost among all the network architectures. Actually, \emph{TimesNet-LE} with the linear head can compete against architectures with the nonlinear head in other classes, which shows excellent property of embedding provided by \emph{TimesNet-LE}. At last, the lengths of output features are being considered. For \emph{TimesNet} with linear heads, features with length of 32 or 256 do not show much difference in performance. For \emph{TimesNet-LE}, features with length of 32 are used with the linear head or the nonlinear head. Even though the feature length is decreased for \emph{TimesNet-LE}, its performance is still improved, which shows that we can use much more compact features for an efficient backbone.

\paragraph{Uncorrelated noise.} The noise originates from the terminal side are uncorrelated and can be viewed as ideal Gaussian white noise. In this condition, all network architectures tend to improve when signal quality gets better, with some fluctuations due to randomness in the training process. In Fig. \ref{fig:lux-perf-tc-uncorr} and Fig. \ref{fig:lux-perf-ratio-uncorr}, it can be seen that, for the same class of networks, the nonlinear head still shows better performance than the linear head. Besides, architectures with the backbone still perform better than architectures without the backbone when they share the same heads, which shows the advantage of features extracted from \emph{TimesNet} or \emph{TimesNet-LE}. Among all the network architectures, \emph{TimesNet-LE} with the nonlinear head achieves the best results again, although it is based on features only with length of 32.

By comparison between the cases with \emph{correlated} and \emph{uncorrelated} noise, it can be found that the performance measures are indeed similar at high SNR; when SNR gets lower, \emph{correlated} noise has more significant effects on the performance of neural networks than \emph{uncorrelated} noise. From the perspective of information theory, in each waveform, sampling points with \emph{uncorrelated} noise (Gaussian white noise) are independent and contain the maximal information about the ground-truth parameters, whereas sampling points with \emph{correlated} noise (noise after a low-pass filter) are not independent and have partial information overlap. Hence, under the same SNR, neural network models can give the best achievable results if the noise is \emph{uncorrelated}.

\begin{figure}[htb]
	\centering
	\includegraphics[width=0.98\textwidth]{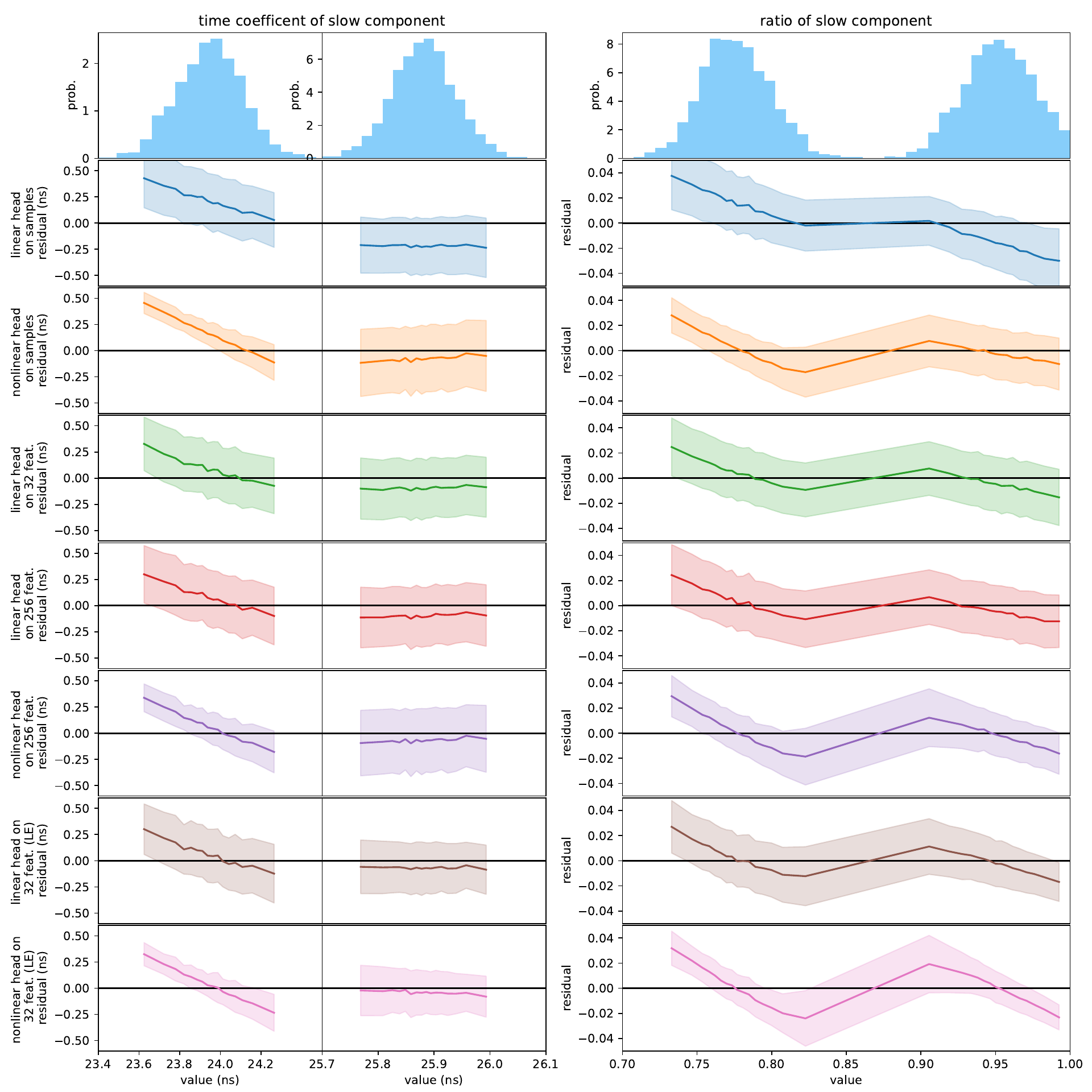}
	\caption{The distributions of ground-truth parameters, and the error bands of residuals segmented on percentiles of ground-truth parameters with \emph{correlated} noise. Different regression models are evaluated in this figure.}
	\label{fig:lux-corr-segment}
\end{figure}

\begin{figure}[htb]
	\centering
	\includegraphics[width=0.98\textwidth]{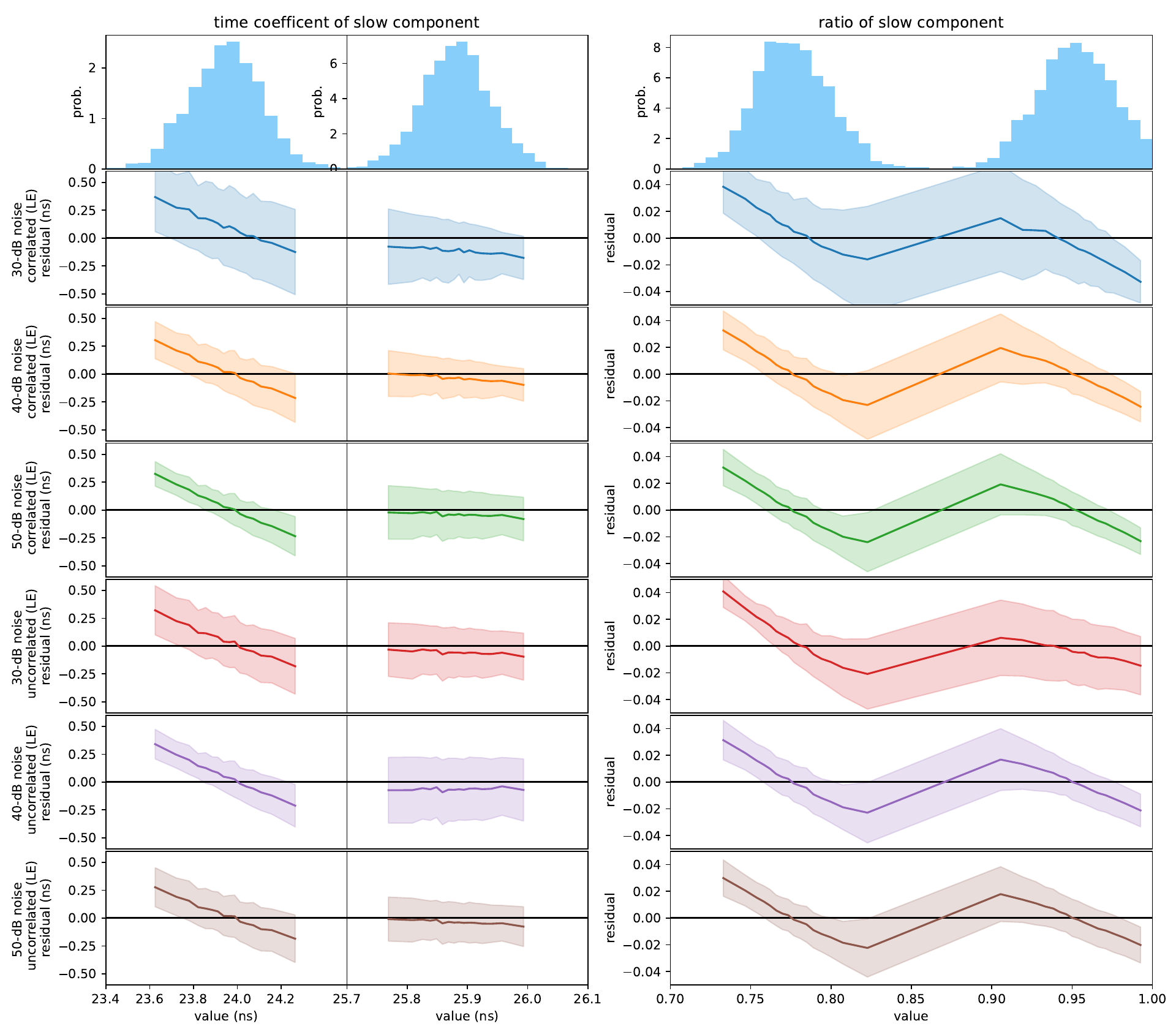}
	\caption{The distributions of ground-truth parameters, and the error bands of residuals segmented on percentiles of ground-truth parameters for \emph{TimesNet-LE} with the nonlinear head. Different noise conditions are evaluated in this figure.}
	\label{fig:timesnet-le-segment}
\end{figure}

To dig into the performance of different network architectures, we have plotted the histograms on distributions of ground-truth parameters. Then we segment the predictions by their corresponding ground-truth parameters on certain percentiles, calculate the mean and standard deviation on each segment, and gather results into error bands. The histograms of ground-truth parameters and error bands of different network architectures at SNR of 50 dB are shown in Fig. \ref{fig:lux-corr-segment}. By examining the figure, it can be found that the errors of neural networks can be decomposed into two parts: a \emph{systematic} part which manifests as drift away from the central line, and a \emph{random} part which manifests as the width of the error band. By increasing the network complexity by using more nonlinear connections, the \emph{systematic} part can be significantly reduced; however, the \emph{random} part does not fully obey the rule. Improving the \emph{random} part relies on better structures in network architectures. It can be found that the last line (\emph{TimesNet-LE} with the nonlinear head) has the least errors both in the \emph{systematic} part and in the \emph{random} part, which shows the fitness of the proposed method for feature extraction of scintillation signals.

Besides, we use the same segmentation method to analyze \emph{TimesNet-LE} with the nonlinear head under different noise conditions. This is illustrated in Fig. \ref{fig:timesnet-le-segment}. The rows from 2 to 4 show the cases with \emph{correlated} noise, and rows from 5 to 7 show the cases with \emph{uncorrelated} noise. It can be validated in this figure that \emph{correlated} noise has more significant effects on regression performance than \emph{uncorrelated} noise. With higher SNR, \emph{TimesNet-LE} with the nonlinear head tends to converge to two modalities and has less deviations from the central lines.

As a final comment, the above studies based on the simulation environment can display the advantage of the proposed method from some perspective, but not all its potentials. In the following section, we will show additional results based on the experiment to more comprehensively evaluate the proposed method.

\subsection{NICA/MPD calorimeter} 

\subsubsection{Experiment setup}

\begin{figure}[htb]
	\centering
	\includegraphics[width=0.8\textwidth]{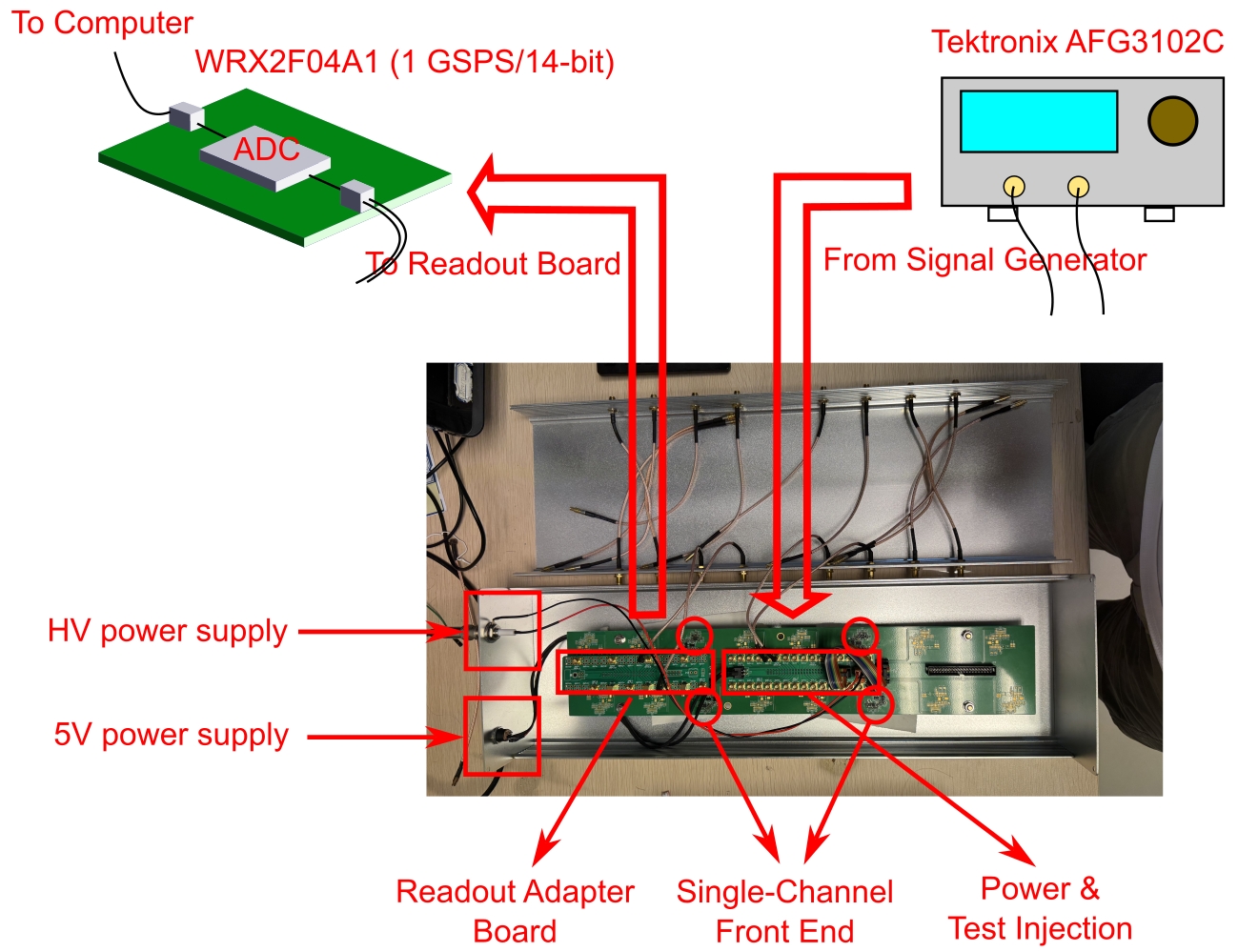}
	\caption{Diagram and photograph of the experimental system. (HV: High Voltage, ADC: Analog-to-Digital Converter, GSPS: Giga Samples Per Second)}
	\label{fig:exp-system-ecal}
\end{figure}

\begin{figure}[htb]
	\centering
	\includegraphics[width=0.98\textwidth]{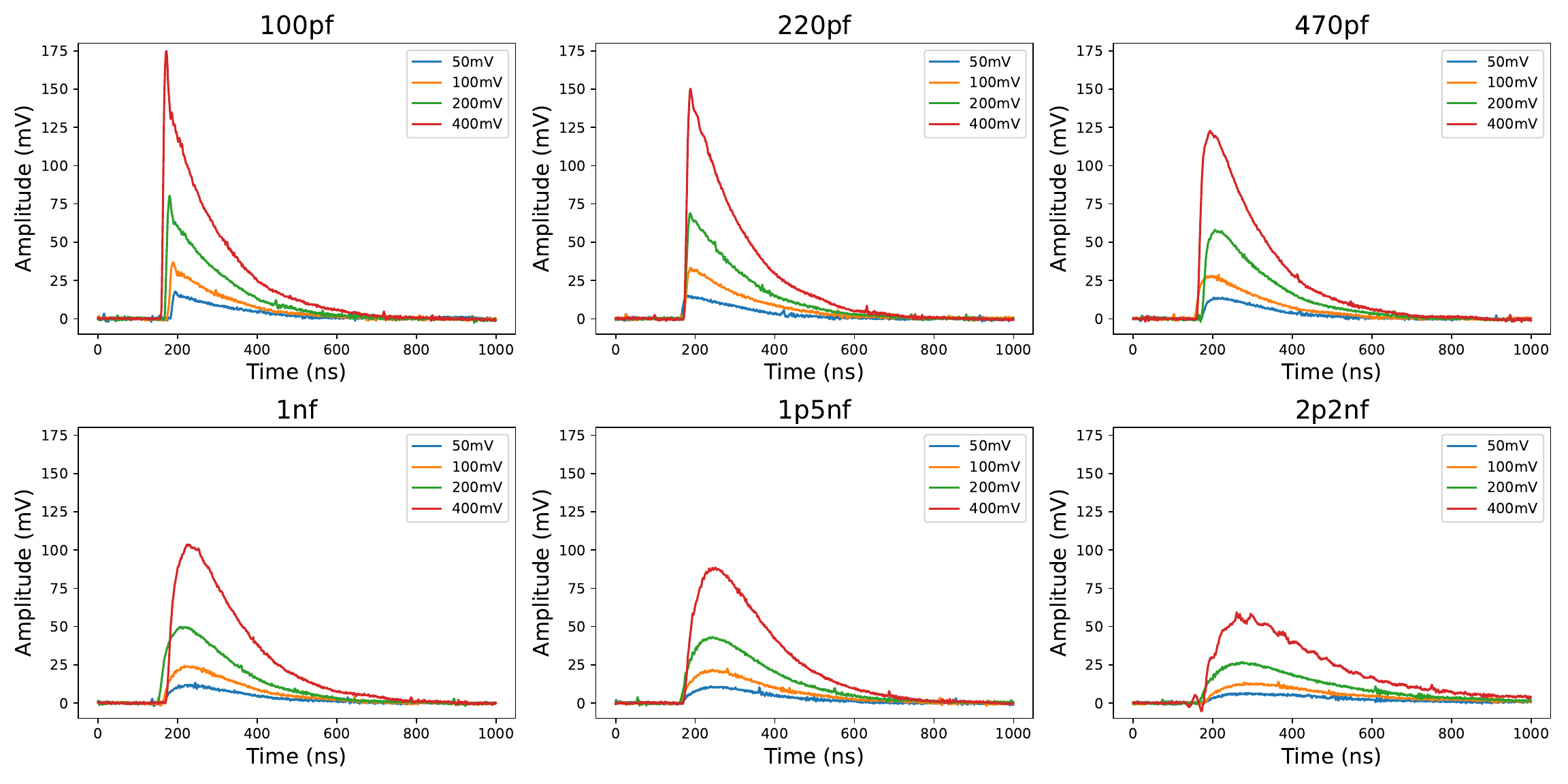}
	\caption{Experimental data emulating scintillation signals with fast electronics. In each sub-figure, we plot an example waveform for each injection voltage at the front-end capacitance indicated by the sub-title.}
	\label{fig:data-viz-ecal}
\end{figure}

In the experiment, we use the fast electronics to emulate variations of scintillation signals. The fast electronics is taken from a \emph{shashlik} electromagnetic calorimeter \cite{NICA-MPD-ECAL-TDR} (with alternate plastic scintillators and lead absorbers) for the Multi-Purpose Detector (MPD) in the Nuclotron-based Ion Collider fAcility (NICA) \cite{NICA-MPD-CDR}. The detector system used in the experiment is shown in Fig. \ref{fig:exp-system-ecal}. Square pulses with preset amplitudes of voltage are generated by the arbitrary function signal generator (Tektronix AFG3102C). These pulses are injected into the fast electronics through an adapter board for power \& test injection. The electronics is surrounded by a shielding box to avoid electromagnetic interference. Test pulses reach the single-channel front ends with preset capacitance to produce response signals. Then the response signals are wired to another adapter board for readout and connected to a custom application-specific integrated circuit (WRX2F04A1 with 1-GSPS rate and 14-bit precision) for analog-to-digital conversion. Digitized signals are finally sent to computers for storage and analysis.

The NICA/MPD calorimeter is able to detect different types of particles (electrons, photons, charged hadrons, muons, etc.). Due to the diversity of shower development, these particles deposit partial or whole energy in the calorimeter and generate varied shapes of electronic pulses. By adjusting the injection voltage from the signal generator, and replacing the front-end capacitance in single-channel circuits, we can change the appearances of response signals to reflect these variations. This is shown in Fig. \ref{fig:data-viz-ecal}. In each sub-figure, we plot four values of injection voltage from 50 mV to 400 mV at the specific front-end capacitance indicated by the sub-title. The front-end capacitance is varied in the range between 100 pF to 2.2 nF. With larger capacitance, the pulse appears to have longer rise time, and more noise charge is induced as the charge is the product of capacitance and voltage.

The main goal of the experiment is to distinguish signals with different values of injection voltage and front-end capacitance. Since there are 24 combinations of the two parameters in total, this is a multi-class classification problem which covers both energy estimation and pulse shape discrimination. We add a minimum level of noise (0.001 relative to the signal amplitude, equivalent to 60 dB) to experimental data for regularization purposes. No other filtering or pre-processing operations are conducted on the waveform samples.

\subsubsection{Results}

In the experiment, the 1-GSPS raw waveform is downsampled with a ratio 5:1, and a section of 384 samples is randomly selected on the downsampled waveform. This gives an input time series with 5-ns period (or equivalent 200-MSPS rate) and 384 elements. We have gathered approximately 4143 events for each condition with specific injection voltage and front-end capacitance, forming a dataset with 99,438 examples in total. Then the dataset is divided into a training set with 69,600 examples, a validation set with 9942 examples and a test set with 19,896 examples.

\begin{table}[htb]
	\centering
	\caption{Numbers of trainable parameters in different network architectures in the experiment.}
	\label{tab:param-exp}
	\small
	\begin{tabularx}{\textwidth}{bqu}
		\hline
		\textbf{network architecture} & \textbf{\#param (encoder)} & \textbf{\#param (head)} \\
		\hline
		linear head on 256 feat. (\emph{TimesNet}) & 684.9k & 6.2k \\
		nonlinear head on 256 feat. (\emph{TimesNet}) & 684.9k & 36.0k \\
		linear head on samples & None & 9.2k \\
		nonlinear head on samples & None & 52.4k \\
		linear head on 32 feat. (\emph{TimesNet-LE}) & 31.5k & 0.8k \\
		nonlinear head on 32 feat. (\emph{TimesNet-LE}) & 31.5k & 1.8k \\
		\hline
	\end{tabularx}
\end{table}

We totally consider 6 network architectures grouped by 3 classes, listed in table \ref{tab:param-exp} with corresponding numbers of trainable parameters in the encoder and in the head. It can be seen that \emph{TimesNet-LE} is still much more compact than \emph{TimesNet}. These network architectures are very similar to ones in simulation, except for the input dimension and the structure of the classification head. More details about the configuration and network architectures can be found in Appendix \ref{sec:detail-net-arch}.

\begin{table}[htb]
	\centering
	\begin{ThreePartTable}
		\caption{Top-1 accuracy to classify the injection voltage, the front-end capacitance and the overall cross conditions.}
		\label{tab:res-exp}
		\footnotesize
		\begin{tabularx}{\textwidth}{bkkm}
			\hline
			\textbf{method} & \textbf{vol. acc. (\%)} & \textbf{cap. acc. (\%)} & \textbf{overall (\%)} \\
			\hline
			Nearest Centroid & 95.90 & 72.69 & --\tnote{a} \\
			Random Forest & 95.90 & 82.23 & 92.95 \\
			\hline
			linear head on 256 feat. (\emph{TimesNet}) & 65.11 & 35.80 & 29.42 \\
			nonlinear head on 256 feat. (\emph{TimesNet}) & 82.20 & 38.98 & 36.48 \\
			linear head on samples & 95.90 & 59.70 & 57.43 \\
			nonlinear head on samples & 95.90 & 78.77 & 75.73 \\
			linear head on 256 feat. (\emph{TimesNet}, finetuned) & 95.83 & 89.12 & 85.80 \\
			nonlinear head on 256 feat. (\emph{TimesNet}, finetuned) & 95.92 & 81.84 & 79.29 \\
			linear head on 32 feat. (\emph{TimesNet-LE}) & 95.85 & 97.99 & 94.00 \\
			nonlinear head on 32 feat. (\emph{TimesNet-LE}) & 95.90 & 98.92 & 94.94 \\
			\hline
		\end{tabularx}
		\begin{tablenotes}
			\item[a] We do not include the overall performance of nearest centroid because it is very poor and not representative when considering rise time and waveform integral as input features at the same time.
		\end{tablenotes}
	\end{ThreePartTable}
\end{table}

\begin{figure}[h!]
	\centering
	\includegraphics[width=0.8\textwidth]{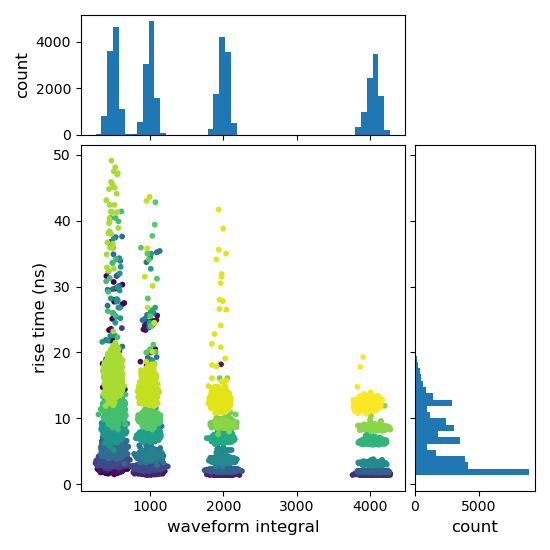}
	\caption{Extraction of rise time and waveform integral from examples in the test dataset for use in conventional methods. Each color in the scatter plot represents examples with a specific injection voltage and front-end capacitance.}
	\label{fig:feat-scatter-hist}
\end{figure}

\begin{figure}[h!]
	\centering
	\includegraphics[width=0.98\textwidth]{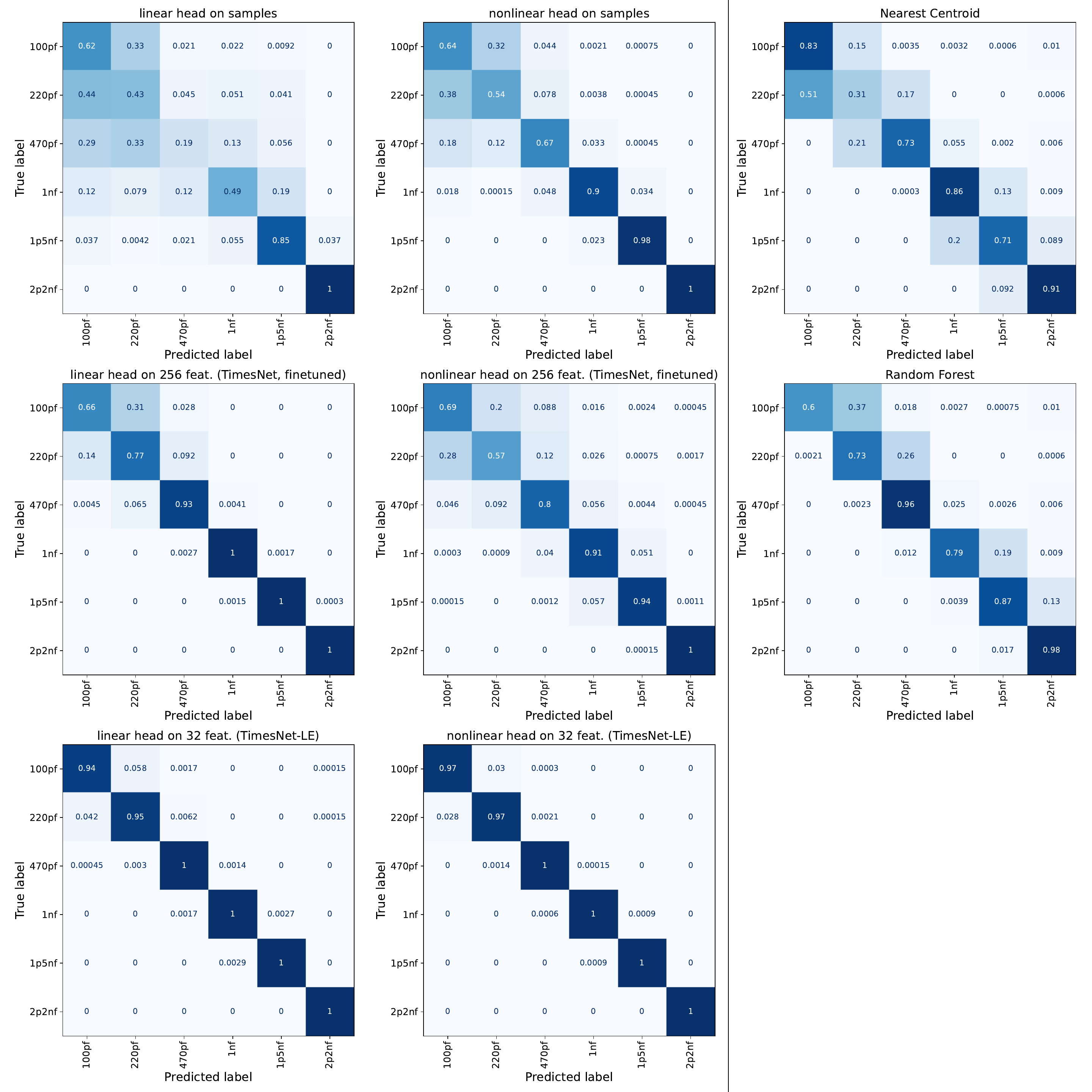}
	\caption{Confusion matrices of different methods identifying the front-end capacitance.}
	\label{fig:ecal-cm-cap}
\end{figure}

\begin{figure}[h!]
	\centering
	\includegraphics[width=0.98\textwidth]{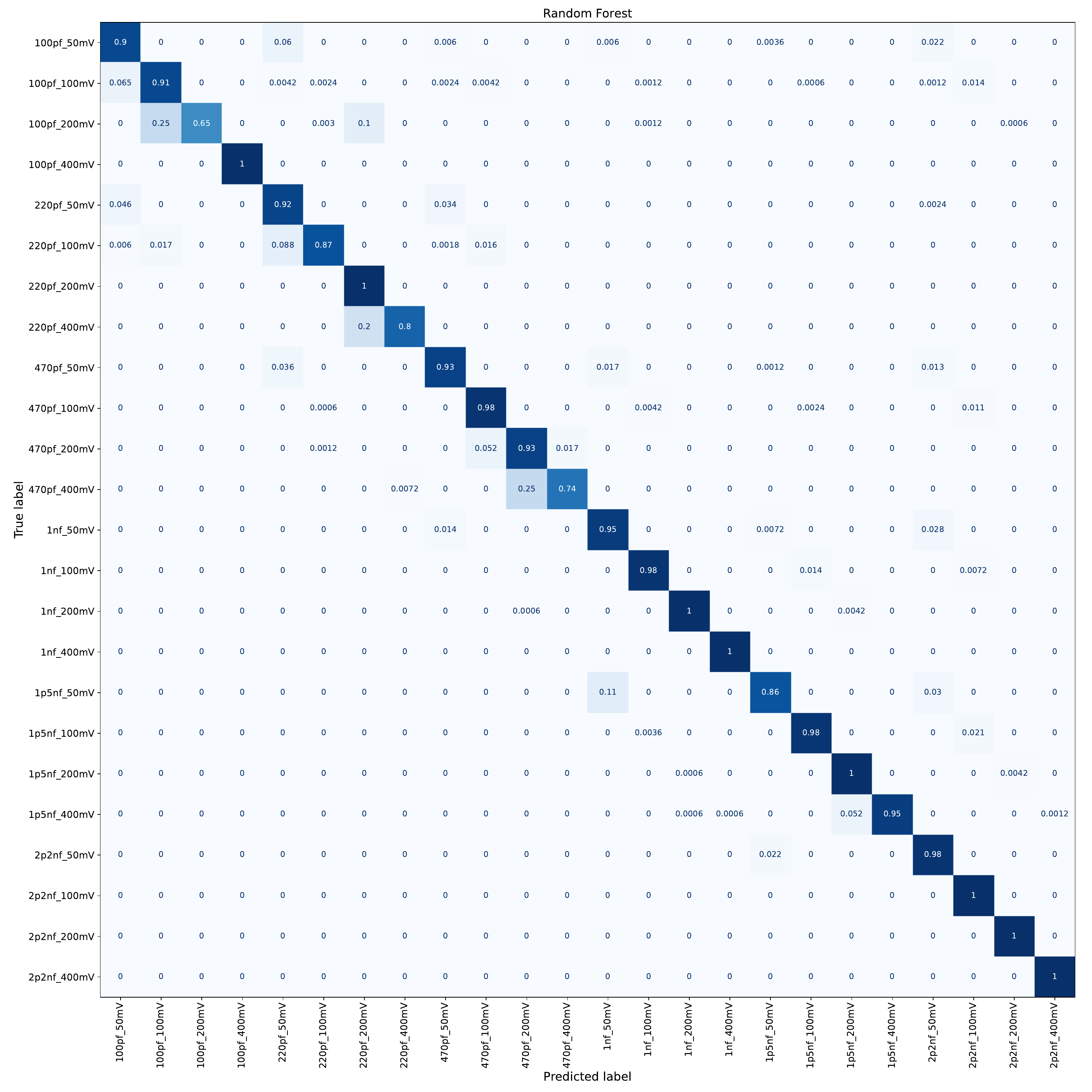}
	\caption{The confusion matrix of random forest identifying the cross conditions of injection voltage and front-end capacitance.}
	\label{fig:ecal-cm-rf}
\end{figure}

\begin{figure}[h!]
	\centering
	\includegraphics[width=0.98\textwidth]{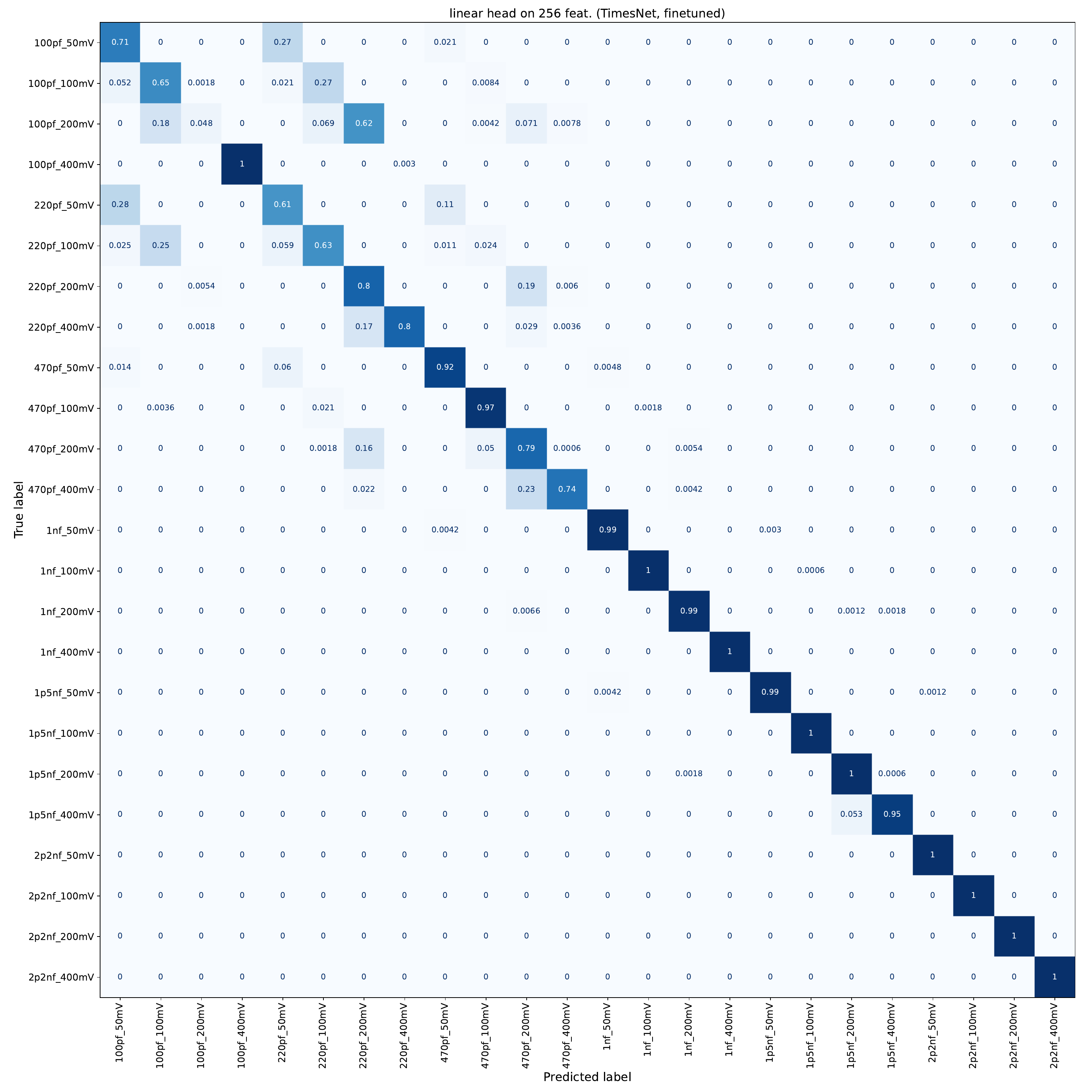}
	\caption{The confusion matrix of the finetuned \emph{TimesNet} model with the linear head identifying the cross conditions of injection voltage and front-end capacitance.}
	\label{fig:ecal-cm-h1-ft}
\end{figure}

\begin{figure}[h!]
	\centering
	\includegraphics[width=0.98\textwidth]{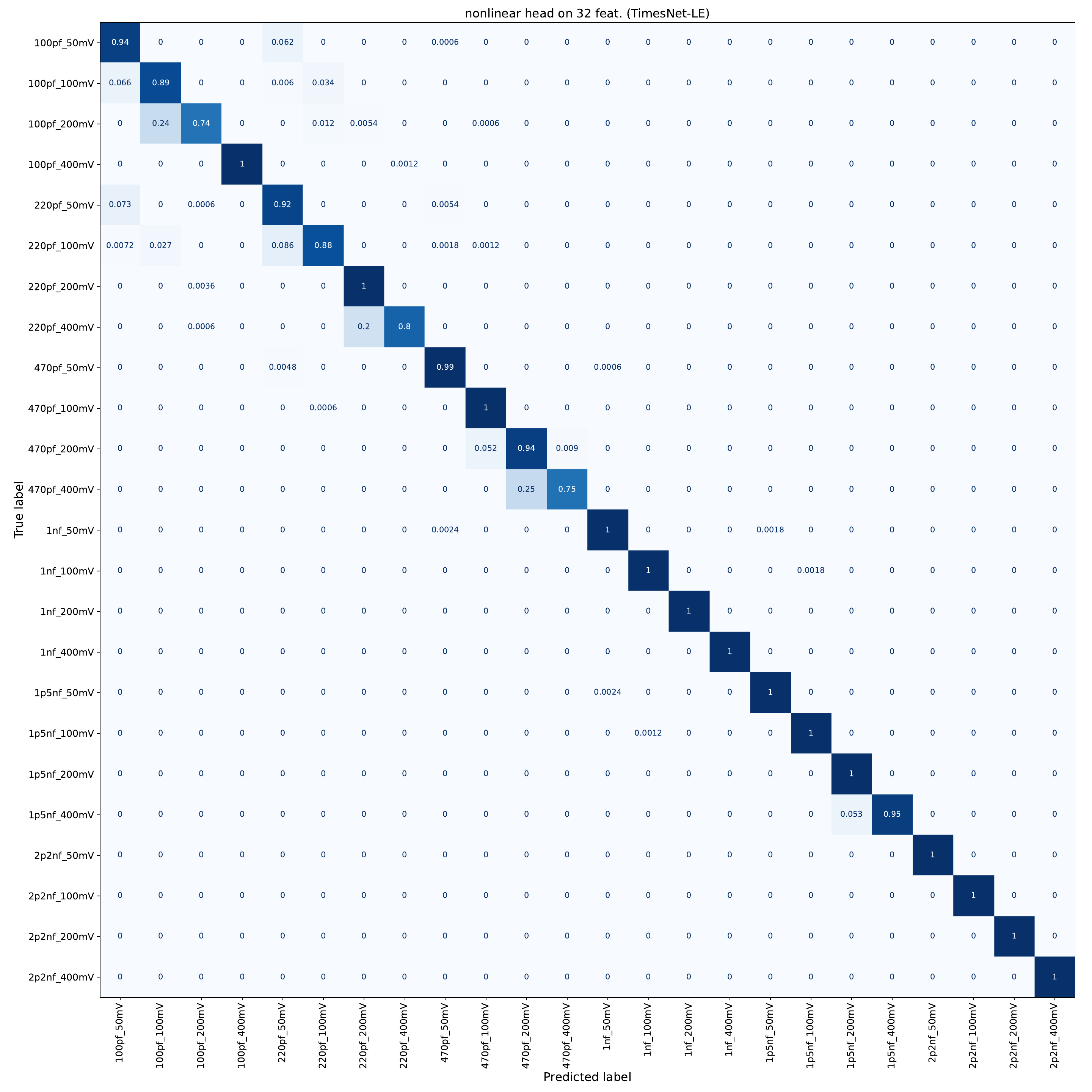}
	\caption{The confusion matrix of the \emph{TimesNet-LE} model with the nonlinear head identifying the cross conditions of injection voltage and front-end capacitance.}
	\label{fig:ecal-cm-le-f32-h1}
\end{figure}

The neural network models are trained to predict 1 out of 24 classes (the cross product of 4 values of injection voltage and 6 values of front-end capacitance) in total. To judge the performance, values of injection voltage and front-end capacitance are extracted from the prediction. Then the accuracy of each factor is calculated separately with related ground-truth labels. Besides, the accuracy as regards the cross product of two factors is also calculated and analyzed.

In comparison, we also consider two conventional machine learning methods: \emph{nearest centroid} and \emph{random forest} (in Scikit-learn Python package \cite{scikit-learn}). In order to apply these two methods, we first extract rise time and waveform integral from each example in the test dataset. To improve the precision of rise time, we interpolate the signals with a ratio of 1:10 to find the 10\%/90\% crossing points (relative to the peak amplitude) at the rise edge. The results of feature extraction are shown in Fig. \ref{fig:feat-scatter-hist}. It can be seen that waveform integral reflecting the injection voltage has separate distributions, whereas rise time at different values of the injection voltage has large deviations. The wide spread of rise time at low injection voltage is due to the difficulty of exact peak finding and crossing-point determination at low-energy and noisy conditions.

The top-1 accuracy of these classification tasks by different methods is shown in table \ref{tab:res-exp}. For nearest centroid and random forest, injection voltage is predicted solely on waveform integral, and front-end capacitance is predicted solely on rise time. In terms of predicting the injection voltage, all methods are able to classify different values with accuracy well above random guesses. Among them, \emph{TimesNet} models without finetuning give poorer results than other methods which show very similar good performance. This indicates the data distribution in the experiment is too challenging for SimSiam (section \ref{sec:rep-learn}) to generate highly informative embedding. Hence, we finetune the backbones and prediction heads of \emph{TimesNet} models as a separate training stage. After finetuning, the performance of \emph{TimesNet} models has been significantly improved.

In terms of predicting the front-end capacitance, different methods show varied scores. For conventional methods, random forest (82.23\%) is better than nearest centroid (72.69\%). Network architectures directly applying neural networks on samples (59.70\%, 78.77\%) are not as good as those with finetuned \emph{TimesNet} backbones (89.12\%, 81.84\%), but network architectures with \emph{TimesNet-LE} backbones (97.99\%, 98.92\%) can be even better. In total, the results indicate that \emph{TimesNet-LE} with reduced model parameters and improved inner structures is an effective feature extractor for this classification task, and can be a justified choice for performance enhancement.

To analyze the reasons why different methods give different performance, we compute the confusion matrices for front-end capacitance by each method. This is shown in Fig. \ref{fig:ecal-cm-cap}. For conventional methods, the ambiguity exists for nearly all values of front-end capacitance. Networks on samples show ambiguity especially for small values of front-end capacitance, because the fast rising edges in these conditions are not utilized properly by the methods. The ambiguity for small values still persists for finetuned networks with \emph{TimesNet} backbones, but the ambiguous regions become narrower. In comparison, networks with \emph{TimesNet-LE} backbones have good classification ability not only in regions with large values, but also in regions reaching the smallest values. By visualization of prediction results, we can explain why \emph{TimesNet-LE} with the nonlinear head scores the best among all the classification methods.

Finally, we select random forest and two best-performing network architectures and plot the confusion matrices for classifying the cross conditions of injection voltage and front-end capacitance. These are shown in Fig. \ref{fig:ecal-cm-rf} (for applying random forest), Fig. \ref{fig:ecal-cm-h1-ft} (for applying the linear head on the \emph{TimesNet} backbone), and Fig. \ref{fig:ecal-cm-le-f32-h1} (for applying the nonlinear head on the \emph{TimesNet-LE} backbone). For random forest, there are light-colored squares scattering in a wide area in Fig. \ref{fig:ecal-cm-rf}, which shows the errors are random and not restricted to a few conditions. For the linear head on the \emph{TimesNet} backbone, there are several light-colored lines parallel to the main diagonal line in Fig. \ref{fig:ecal-cm-h1-ft}, which corresponds to the misclassified examples into the wrong classes of the front-end capacitance, and this is the main source of decreased overall accuracy for this network architecture. In contrast, the confusion matrix in Fig. \ref{fig:ecal-cm-le-f32-h1} is the cleanest with few misclassified examples in the upper left region, and almost all examples fall into the main diagonal elements. The detailed visualization demonstrate the fitness of \emph{TimesNet-LE} as a representation learning backbone for multi-class classification.

\section{Discussion}

Throughout the case studies, we illustrate how to use spectrum-inspired temporal neural networks as representation learning backbones to improve the performance of complex tasks for scintillation signals. Here we present a discussion mainly on the applicability of the proposed method.

In the first case study, several network architectures are studied on a multi-target (time coefficient and ratio of slow components) and multimodal (ER and NR events) regression task. To generate raw waveforms, we consider scintillation emission, optical process, electronics and noise, making the analytical form of the signals very realistic but also complicated. Directly applying traditional methods, such as curve fitting, will face challenges from numeric issues to ill-conditioning. Besides, when data points are correlated by low-pass filtering of the Gaussian white noise, effective curve fitting accounting for the covariance matrix of data points is impractical and compute-intensive. In comparison, neural networks (like \emph{TimesNet-LE}) approximate the regression task with universal functional mapping between waveform features and regression targets. Although neural networks are agnostic to the analytic form of the signals, their strength of learning from limited data examples gives them a unique advantage. Specially, the network architecture proposed in this paper can utilize temporal and spectral features more efficiently, demonstrating it as a good candidate for similar regression tasks.

In the second case study, network architectures, as well as conventional machine learning methods, are applied to a multi-class (multiple pulse shapes and energies) classification task. Two features (rise time and waveform integral) are extracted from the examples to facilitate conventional methods, such as nearest centroid and random forest (similar to boosted decision tree). It can be seen that advanced network architectures (like \emph{TimesNet-LE}) give the best overall performance, but results from random forest are also competitive. However, the high accuracy of random forest heavily relies on the feature extraction process, with interpolation, peak finding and crossing-point determination as most important steps. Though these steps can be realized on high-end computers, the iterative nature makes them hard to be pipelined and limits their application for online inference. On the other hand, assisted by modern libraries of neural network synthesis for hardware \cite{Khoda_2023}, compact network architectures can be easily converted to digital logic for low-latency and highly efficient hardware deployment.

It should be noted that the proposed method is generally applicable to nuclear radiation detectors, not restricted to scintillation detectors. For example, in pulse sampling-based fluorescence lifetime imaging \cite{Zhou:21}, \emph{TimesNet-LE} can be used to estimate fluorescence lifetime values from waveforms coming out of avalanche photodiodes or microchannel plate photomultiplier tubes. In three-dimensional position-sensitive detection \cite{10857362}, \emph{TimesNet-LE} can be used for multi-channel feature extraction with semiconductor-coupled silicon detectors. In data acquisition systems of high-energy physics experiments \cite{TOPKO2022166680}, \emph{TimesNet-LE} can be used for on-detector data pre-processing with double-sided silicon strip detectors.

Finally, features from the same network architecture have the potential to serve some more sophisticated tasks, like uncertainty estimation \cite{Ai_2022} or covariance matrix estimation \cite{DBLP:conf/nips/MishkinKNSK18}, if possible with the signal quality and sampling rate.

\section{Conclusion}

In this paper, we propose an innovative network architecture, specially customized for scintillation pulse characterization, based on previous works on time series analysis. The core idea is to fully exploit the underlying information structure in scintillation pulses by more interpretable data embedding and more emphasis on 0-frequency and low-frequency components in the spectrum. The designed architecture, named \emph{TimesNet-LE}, performs as an excellent representation learning backbone and surpasses its reference model and densely connected models without backbones. Simulation and experimental studies demonstrate the superior performance of the network architecture even with much fewer trainable parameters than its reference model, in a wide area of tasks from multimodal regression to multi-class classification. Two conventional machine learning methods (nearest centroid and random forest) are also compared and discussed.

There are some future directions worthy of exploration. For example, to validate the network architecture in a multi-task setting with mixed prediction heads sharing a common backbone, or to implement the lightweight model on field programmable gate arrays for fast online inference. We sincerely hope the proposed network architecture can benefit related studies in the community and show its feasibility in more real-world experiments.

\backmatter

\bmhead{Acknowledgements}

This research is supported in part by the National Natural Science Foundation of China (under Grant No. 12405228), in part by the National Key Research and Development Program of China (under Grant Nos. 2020YFE0202001, 2024YFF0726201), in part by the Fundamental Research Funds for the Central Universities (under Grant No. CCNU23XJ013) and in part by the China Postdoctoral Science Foundation (under Grant No. 2023M731244).

\section*{Declarations}

\bmhead{Competing Interests}

The authors declare that they have no competing interests.

\bmhead{Data availability statement}

The manuscript has associated data in a data repository.

\begin{appendices}

\section{Details of simulation}
\label{sec:detail-sim}

Simulation of the LUX detector starts with the scintillation emission comprising a fast component and a slow component:

\begin{equation} \label{equ:scint-emi}
	f_1(t) = \left( \frac{A_1}{1 + A_1} \cdot \frac{1}{\tau_f} e^{-t/\tau_f} + \frac{1}{1 + A_1} \cdot \frac{1}{\tau_s} e^{-t/\tau_s} \right) u(t)
\end{equation}

\noindent where $\tau_f, \tau_s$ are time coefficients of the fast and slow components, and $A_1$ is the ratio between fast and slow components. $u(t)$ is the step function taking $1$ when $t \geq 0$ and $0$ elsewhere.

Emitted photons travel to sensors through an optical process featuring a direct path and a weighted decay path due to reflection and scattering:

\begin{equation} \label{equ:optic-proc}
	f_2(t) = A_2 \delta (t) + (1 - A_2) \left( \frac{B_a}{\tau_a} e^{-t/\tau_a} + \frac{B_b}{\tau_b} e^{-t/\tau_b} \right) u(t)
\end{equation}

\noindent where $\tau_a, \tau_b$ are time coefficients for the weighted decay path, $B_a, B_b$ are corresponding ratios satisfying $B_a + B_b = 1$, and $A_2$ is the ratio of the direct path. $\delta (t)$ is the Dirac function which takes non-zero value only at $t = 0$ and has integral of $1$.

Finally, the photons are transduced into scintillation signals by the sensors, and a second-order low-pass filter is used to simulate the signal generation process:

\begin{equation} \label{equ:filter}
	f_3(t) = K \cdot \frac{t}{\tau_c} e^{-t/\tau_c} u(t)
\end{equation}

\noindent where $\tau_c$ is the time coefficient of the low-pass filter, and $K$ is an amplitude variable. $\tau_c$ is kept as 40 ns to generate a low-pass filter with 25 MHz critical frequency, and $K$ is sampled from a uniform distribution between $1$ and $2$.

The pulse seen at the sampling side is the convolution integral of the above three signals:

\begin{align}
	f(t) &= f_1(t) * f_2(t) * f_3(t) \nonumber \\
	     &= \int_{-\infty}^{+\infty}\int_{-\infty}^{+\infty}f_1(\tau_2)f_2(\tau_1 - \tau_2)f_3(t - \tau_1)\mathrm{d}\tau_2\mathrm{d}\tau_1
\end{align}

The mean and standard deviation used to generate ER and NR events are listed in table \ref{tab:er} and table \ref{tab:nr}. It should be noted that examples with out-of-range values (e.g. too small values of $\tau_f$) are discarded.

\begin{table}[htb]
	\centering
	\caption{Mean and standard deviation used to generate ER events. We use hyphens to denote no standard deviations are considered. The units for time coefficients are nanoseconds.}
	\label{tab:er}
	\small
	\begin{tabularx}{\textwidth}{XXXXXXXX}
		\hline
		 & $A_1$ & $\tau_f$ & $\tau_s$ & $A_2$ & $B_a$ & $\tau_a$ & $\tau_b$ \\
		\hline
		mean & 0.04389 & 2.305 & 25.89 & 0.0574 & 1.062 & 11.1 & 2.70 \\
		std. & 0.02961 & 1.324 & 0.06 & -- & -- & -- & -- \\
		\hline
	\end{tabularx}
\end{table}

\begin{table}[htb]
	\centering
	\caption{Mean and standard deviation used to generate NR events. We use hyphens to denote no standard deviations are considered. The units for time coefficients are nanoseconds.}
	\label{tab:nr}
	\small
	\begin{tabularx}{\textwidth}{XXXXXXXX}
		\hline
		& $A_1$ & $\tau_f$ & $\tau_s$ & $A_2$ & $B_a$ & $\tau_a$ & $\tau_b$ \\
		\hline
		mean & 0.28245 & 2.305 & 23.97 & 0.0574 & 1.062 & 11.1 & 2.70 \\
		std. & 0.04035 & 1.324 & 0.17 & -- & -- & -- & -- \\
		\hline
	\end{tabularx}
\end{table}

To generate correlated noise as a Gaussian process, we compute the frequency response of the low-pass filter, and use the frequency response to generate the auto-correlation function of the noise:

\begin{align}
	H_3(f) &= \mathscr{F}\left[ f_3(t) \right] \nonumber \\
	P_o(f) &= \left| H_3(f) \right|^2 P_i(f) \nonumber \\
	R_o(\tau) &= \mathscr{F}^{-1}\left[ P_o(f) \right] = \mathscr{F}^{-1}\left[ \left| \mathscr{F}\left[ f_3(t) \right] \right|^2 P_i(f) \right] \label{equ:noise}
\end{align}

\noindent where $\mathscr{F}[\cdot]$ and $\mathscr{F}^{-1}[\cdot]$ denote Fourier and inverse Fourier transforms. Then the covariance matrix of the noise is generated by sampling the auto-correlation function. Finally, the correlated noise is sampled from a multivariate Gaussian distribution with zero mean and this covariance matrix.

On the other hand, the uncorrelated noise is generated by sampling from a multivariate Gaussian distribution with zero mean and a diagonal covariance matrix.

\section{Details of configuration and network architectures}
\label{sec:detail-net-arch}

Throughout the work, we use PyTorch \cite{DBLP:conf/asplos/AnselYHGJVBBBBC24} to implement neural networks. For regression tasks in simulation, we use mean squared error as loss function; for classification tasks in the experiment, we use cross entropy as loss function. For tasks with linear heads, we use stochastic gradient optimization with 0.9 momentum and no weight decay as our optimizer; for tasks with nonlinear heads, we use Adam \cite{DBLP:journals/corr/KingmaB14} with default parameters as our optimizer. The initial learning rate (0.0005$\sim$0.1) and training epochs (20$\sim$80) are adjusted in each training process for fast and reliable convergence. The batch size is set to 512 to maximize hardware utilization. We use two NVIDIA GeForce RTX 4060 Ti graphic cards with 2$\times$16 GB video memory to accelerate training.

\subsection{Networks in simulation}
\label{sec:network-sim}

\paragraph{TimesNet.} The input time series with 256 elements is transformed by data embedding, including a constant positional embedding made up of sine and cosine functions and a trainable token embedding using 1D convolution to keep the length and increase the channels from 1 to 16. Two blocks of multi-period convolution (see section \ref{sec:multi-period-conv}) follow to construct the body of the network. In each block, multi-period convolution is performed on top 5 frequencies revealed by the FFT spectrum. For each frequency/period, a combination of convolution kernels, with kernel sizes of (1, 1), (3, 3), (5, 5), (7, 7), (9, 9) and (11, 11), is used for 2D convolution, with appropriate padding and unit stride to keep the tensor shape unchanged. The 2D convolution has a structure with mirroring convolution layers to first increase the channels to 32 and then reduce to 16. A layer of Gaussian Error Linear Units (GELU) is used at the middle as the activation function. Finally, the output of the last multi-period convolution block is projected to a fixed-length (32 or 256) feature vector by a linear layer.

\paragraph{TimesNet-LE.} The data embedding is bypassed, and the input time series with 256 elements directly enters the only one block of multi-period convolution. Multi-period convolution is performed on top 5 frequencies revealed by the FFT spectrum. For each frequency/period, a combination of convolution kernels, with kernel sizes of (1, 1), (3, 3) and (5, 5), is used for 2D convolution, while keeping the tensor shape unchanged. The mirroring convolution layers first increase channels from 1 to 32 and then reduce to 16, with GELU in between. The output is projected to a fixed-length (32) feature vector by a linear layer.

\paragraph{Heads.} For linear heads, a linear layer of multiplication and summation (with no activation function) directly maps input samples or features to 2 regression targets. For nonlinear heads, a densely connected network with one hidden layer of 8 (for 32 features) or 64 (for others) neurons and Rectified Linear Units (ReLU) maps input samples or features to each regression target.

\subsection{Networks in the experiment}

\paragraph{TimesNet.} Similar to \emph{TimesNet} in Appendix \ref{sec:network-sim}. The number of elements in the input time series is changed from 256 to 384, and tensor shapes in intermediate layers are changed accordingly.

\paragraph{TimesNet-LE.} Similar to \emph{TimesNet-LE} in Appendix \ref{sec:network-sim}. The number of elements in the input time series is changed from 256 to 384, and tensor shapes in intermediate layers are changed accordingly.

\paragraph{Heads.} For linear heads, a linear layer directly maps input samples or features to a vector of 24 logits, with each element representing a class. For nonlinear heads, a densely connected network with one hidden layer of 32 (for 32 features) or 128 (for others) neurons and ReLU maps input samples or features to a vector of 24 logits.

\end{appendices}

\bibliography{mybibfile}


\begin{thebibliography}{45}
\ifx \bisbn   \undefined \def \bisbn  #1{ISBN #1}\fi
\ifx \binits  \undefined \def \binits#1{#1}\fi
\ifx \bauthor  \undefined \def \bauthor#1{#1}\fi
\ifx \batitle  \undefined \def \batitle#1{#1}\fi
\ifx \bjtitle  \undefined \def \bjtitle#1{#1}\fi
\ifx \bvolume  \undefined \def \bvolume#1{\textbf{#1}}\fi
\ifx \byear  \undefined \def \byear#1{#1}\fi
\ifx \bissue  \undefined \def \bissue#1{#1}\fi
\ifx \bfpage  \undefined \def \bfpage#1{#1}\fi
\ifx \blpage  \undefined \def \blpage #1{#1}\fi
\ifx \burl  \undefined \def \burl#1{\textsf{#1}}\fi
\ifx \doiurl  \undefined \def \doiurl#1{\url{https://doi.org/#1}}\fi
\ifx \betal  \undefined \def \betal{\textit{et al.}}\fi
\ifx \binstitute  \undefined \def \binstitute#1{#1}\fi
\ifx \binstitutionaled  \undefined \def \binstitutionaled#1{#1}\fi
\ifx \bctitle  \undefined \def \bctitle#1{#1}\fi
\ifx \beditor  \undefined \def \beditor#1{#1}\fi
\ifx \bpublisher  \undefined \def \bpublisher#1{#1}\fi
\ifx \bbtitle  \undefined \def \bbtitle#1{#1}\fi
\ifx \bedition  \undefined \def \bedition#1{#1}\fi
\ifx \bseriesno  \undefined \def \bseriesno#1{#1}\fi
\ifx \blocation  \undefined \def \blocation#1{#1}\fi
\ifx \bsertitle  \undefined \def \bsertitle#1{#1}\fi
\ifx \bsnm \undefined \def \bsnm#1{#1}\fi
\ifx \bsuffix \undefined \def \bsuffix#1{#1}\fi
\ifx \bparticle \undefined \def \bparticle#1{#1}\fi
\ifx \barticle \undefined \def \barticle#1{#1}\fi
\bibcommenthead
\ifx \bconfdate \undefined \def \bconfdate #1{#1}\fi
\ifx \botherref \undefined \def \botherref #1{#1}\fi
\ifx \url \undefined \def \url#1{\textsf{#1}}\fi
\ifx \bchapter \undefined \def \bchapter#1{#1}\fi
\ifx \bbook \undefined \def \bbook#1{#1}\fi
\ifx \bcomment \undefined \def \bcomment#1{#1}\fi
\ifx \oauthor \undefined \def \oauthor#1{#1}\fi
\ifx \citeauthoryear \undefined \def \citeauthoryear#1{#1}\fi
\ifx \endbibitem  \undefined \def \endbibitem {}\fi
\ifx \bconflocation  \undefined \def \bconflocation#1{#1}\fi
\ifx \arxivurl  \undefined \def \arxivurl#1{\textsf{#1}}\fi
\csname PreBibitemsHook\endcsname

\bibitem[\protect\citeauthoryear{Lee et~al.}{2024}]{Lee2024}
\begin{barticle}
\bauthor{\bsnm{Lee}, \binits{S.M.}}, \betal:
\batitle{{Nonproportionality of NaI(Tl) scintillation detector for dark matter
  search experiments}}.
\bjtitle{The European Physical Journal C}
\bvolume{84}(\bissue{5}),
\bfpage{484}
(\byear{2024})
\doiurl{10.1140/epjc/s10052-024-12770-1}
\end{barticle}
\endbibitem

\bibitem[\protect\citeauthoryear{Sun et~al.}{2023}]{Sun2023}
\begin{barticle}
\bauthor{\bsnm{Sun}, \binits{Q.}}, \betal:
\batitle{Dual discrimination of fast neutrons from strong {$\gamma$} noise
  using organic single-crystal scintillator}.
\bjtitle{Matter}
\bvolume{6}(\bissue{1}),
\bfpage{274}--\blpage{284}
(\byear{2023})
\doiurl{10.1016/j.matt.2022.10.010}
\end{barticle}
\endbibitem

\bibitem[\protect\citeauthoryear{Osmanagaoglu and
  Celiktas}{2024}]{Osmanagaoglu2024}
\begin{barticle}
\bauthor{\bsnm{Osmanagaoglu}, \binits{C.Z.}},
\bauthor{\bsnm{Celiktas}, \binits{C.}}:
\batitle{{An MPPC alpha detection performance comparison study by two different
  scintillators}}.
\bjtitle{The European Physical Journal Plus}
\bvolume{139}(\bissue{11}),
\bfpage{1028}
(\byear{2024})
\doiurl{10.1140/epjp/s13360-024-05855-z}
\end{barticle}
\endbibitem

\bibitem[\protect\citeauthoryear{Pandya et~al.}{2025}]{PANDYA2025170177}
\begin{barticle}
\bauthor{\bsnm{Pandya}, \binits{D.}}, \betal:
\batitle{{PSD study of CsI:Tl microcrystals based 3D printed scintillator}}.
\bjtitle{Nuclear Instruments and Methods in Physics Research Section A:
  Accelerators, Spectrometers, Detectors and Associated Equipment}
\bvolume{1072},
\bfpage{170177}
(\byear{2025})
\doiurl{10.1016/j.nima.2024.170177}
\end{barticle}
\endbibitem

\bibitem[\protect\citeauthoryear{Saeidi et~al.}{2025}]{Saeidi2025}
\begin{barticle}
\bauthor{\bsnm{Saeidi}, \binits{Z.}},
\bauthor{\bsnm{Afarideh}, \binits{H.}},
\bauthor{\bsnm{Ghergherehchi}, \binits{M.}}:
\batitle{{Enhanced gamma-ray spectrum transformation: NaI(Tl) scintillator to
  HPGe semiconductor via machine learning}}.
\bjtitle{The European Physical Journal Plus}
\bvolume{140}(\bissue{2}),
\bfpage{113}
(\byear{2025})
\doiurl{10.1140/epjp/s13360-025-06048-y}
\end{barticle}
\endbibitem

\bibitem[\protect\citeauthoryear{Fu et~al.}{2018}]{FU2018410}
\begin{barticle}
\bauthor{\bsnm{Fu}, \binits{C.}},
\bauthor{\bsnm{{Di Fulvio}}, \binits{A.}},
\bauthor{\bsnm{Clarke}, \binits{S.D.}},
\bauthor{\bsnm{Wentzloff}, \binits{D.}},
\bauthor{\bsnm{Pozzi}, \binits{S.A.}},
\bauthor{\bsnm{Kim}, \binits{H.S.}}:
\batitle{Artificial neural network algorithms for pulse shape discrimination
  and recovery of piled-up pulses in organic scintillators}.
\bjtitle{Annals of Nuclear Energy}
\bvolume{120},
\bfpage{410}--\blpage{421}
(\byear{2018})
\doiurl{10.1016/j.anucene.2018.05.054}
\end{barticle}
\endbibitem

\bibitem[\protect\citeauthoryear{Ai et~al.}{2019}]{Ai_2019}
\begin{barticle}
\bauthor{\bsnm{Ai}, \binits{P.}},
\bauthor{\bsnm{Wang}, \binits{D.}},
\bauthor{\bsnm{Huang}, \binits{G.}},
\bauthor{\bsnm{Fang}, \binits{N.}},
\bauthor{\bsnm{Xu}, \binits{D.}},
\bauthor{\bsnm{Zhang}, \binits{F.}}:
\batitle{Timing and characterization of shaped pulses with mhz adcs in a
  detector system: a comparative study and deep learning approach}.
\bjtitle{Journal of Instrumentation}
\bvolume{14}(\bissue{03}),
\bfpage{03002}
(\byear{2019})
\doiurl{10.1088/1748-0221/14/03/P03002}
\end{barticle}
\endbibitem

\bibitem[\protect\citeauthoryear{Aarrestad et~al.}{2021}]{Aarrestad_2021}
\begin{barticle}
\bauthor{\bsnm{Aarrestad}, \binits{T.}}, \betal:
\batitle{{Fast convolutional neural networks on FPGAs with hls4ml}}.
\bjtitle{Machine Learning: Science and Technology}
\bvolume{2}(\bissue{4}),
\bfpage{045015}
(\byear{2021})
\doiurl{10.1088/2632-2153/ac0ea1}
\end{barticle}
\endbibitem

\bibitem[\protect\citeauthoryear{Khoda et~al.}{2023}]{Khoda_2023}
\begin{barticle}
\bauthor{\bsnm{Khoda}, \binits{E.E.}}, \betal:
\batitle{{Ultra-low latency recurrent neural network inference on FPGAs for
  physics applications with hls4ml}}.
\bjtitle{Machine Learning: Science and Technology}
\bvolume{4}(\bissue{2}),
\bfpage{025004}
(\byear{2023})
\doiurl{10.1088/2632-2153/acc0d7}
\end{barticle}
\endbibitem

\bibitem[\protect\citeauthoryear{Ai et~al.}{2023}]{10005128}
\begin{barticle}
\bauthor{\bsnm{Ai}, \binits{P.}},
\bauthor{\bsnm{Deng}, \binits{Z.}},
\bauthor{\bsnm{Wang}, \binits{Y.}},
\bauthor{\bsnm{Gong}, \binits{H.}},
\bauthor{\bsnm{Ran}, \binits{X.}},
\bauthor{\bsnm{Lang}, \binits{Z.}}:
\batitle{{PulseDL-II: A System-on-Chip Neural Network Accelerator for Timing
  and Energy Extraction of Nuclear Detector Signals}}.
\bjtitle{IEEE Transactions on Nuclear Science}
\bvolume{70}(\bissue{6}),
\bfpage{971}--\blpage{978}
(\byear{2023})
\doiurl{10.1109/TNS.2022.3233895}
\end{barticle}
\endbibitem

\bibitem[\protect\citeauthoryear{Doucet et~al.}{2020}]{DOUCET2020161201}
\begin{barticle}
\bauthor{\bsnm{Doucet}, \binits{E.}}, \betal:
\batitle{Machine learning n/{$\gamma$} discrimination in {CLYC} scintillators}.
\bjtitle{Nuclear Instruments and Methods in Physics Research Section A:
  Accelerators, Spectrometers, Detectors and Associated Equipment}
\bvolume{954},
\bfpage{161201}
(\byear{2020})
\doiurl{10.1016/j.nima.2018.09.036}
\end{barticle}
\endbibitem

\bibitem[\protect\citeauthoryear{Zhao et~al.}{2023}]{Zhao_2023}
\begin{barticle}
\bauthor{\bsnm{Zhao}, \binits{K.}},
\bauthor{\bsnm{Feng}, \binits{C.}},
\bauthor{\bsnm{Wang}, \binits{S.}},
\bauthor{\bsnm{Shen}, \binits{Z.}},
\bauthor{\bsnm{Zhang}, \binits{K.}},
\bauthor{\bsnm{Liu}, \binits{S.}}:
\batitle{{n/{$\gamma$} discrimination for CLYC detector using a one-dimensional
  Convolutional Neural Network}}.
\bjtitle{Journal of Instrumentation}
\bvolume{18}(\bissue{01}),
\bfpage{01021}
(\byear{2023})
\doiurl{10.1088/1748-0221/18/01/P01021}
\end{barticle}
\endbibitem

\bibitem[\protect\citeauthoryear{Lee et~al.}{2024}]{LEE2024169638}
\begin{barticle}
\bauthor{\bsnm{Lee}, \binits{S.}}, \betal:
\batitle{{Investigation of neutron/gamma-ray distribution in SiPM-based pulse
  shape discrimination using EJ-276 plastic scintillators}}.
\bjtitle{Nuclear Instruments and Methods in Physics Research Section A:
  Accelerators, Spectrometers, Detectors and Associated Equipment}
\bvolume{1066},
\bfpage{169638}
(\byear{2024})
\doiurl{10.1016/j.nima.2024.169638}
\end{barticle}
\endbibitem

\bibitem[\protect\citeauthoryear{Dutta et~al.}{2023}]{Dutta_2023}
\begin{barticle}
\bauthor{\bsnm{Dutta}, \binits{S.}},
\bauthor{\bsnm{Ghosh}, \binits{S.}},
\bauthor{\bsnm{Bhattacharya}, \binits{S.}},
\bauthor{\bsnm{Saha}, \binits{S.}}:
\batitle{Pulse shape simulation and discrimination using machine learning
  techniques}.
\bjtitle{Journal of Instrumentation}
\bvolume{18}(\bissue{03}),
\bfpage{03038}
(\byear{2023})
\doiurl{10.1088/1748-0221/18/03/P03038}
\end{barticle}
\endbibitem

\bibitem[\protect\citeauthoryear{Jung et~al.}{2023}]{Jung_2023}
\begin{barticle}
\bauthor{\bsnm{Jung}, \binits{K.Y.}}, \betal:
\batitle{Pulse shape discrimination using a convolutional neural network for
  organic liquid scintillator signals}.
\bjtitle{Journal of Instrumentation}
\bvolume{18}(\bissue{03}),
\bfpage{03003}
(\byear{2023})
\doiurl{10.1088/1748-0221/18/03/P03003}
\end{barticle}
\endbibitem

\bibitem[\protect\citeauthoryear{Griffiths et~al.}{2020}]{Griffiths_2020}
\begin{barticle}
\bauthor{\bsnm{Griffiths}, \binits{J.}},
\bauthor{\bsnm{Kleinegesse}, \binits{S.}},
\bauthor{\bsnm{Saunders}, \binits{D.}},
\bauthor{\bsnm{Taylor}, \binits{R.}},
\bauthor{\bsnm{Vacheret}, \binits{A.}}:
\batitle{Pulse shape discrimination and exploration of scintillation signals
  using convolutional neural networks}.
\bjtitle{Machine Learning: Science and Technology}
\bvolume{1}(\bissue{4}),
\bfpage{045022}
(\byear{2020})
\doiurl{10.1088/2632-2153/abb781}
\end{barticle}
\endbibitem

\bibitem[\protect\citeauthoryear{Cheng et~al.}{2024}]{Cheng2024}
\begin{barticle}
\bauthor{\bsnm{Cheng}, \binits{J.}}, \betal:
\batitle{{Pulse shape discrimination technique for diffuse supernova neutrino
  background search with JUNO}}.
\bjtitle{The European Physical Journal C}
\bvolume{84}(\bissue{5}),
\bfpage{482}
(\byear{2024})
\doiurl{10.1140/epjc/s10052-024-12779-6}
\end{barticle}
\endbibitem

\bibitem[\protect\citeauthoryear{Carlini et~al.}{2024}]{CARLINI2024169369}
\begin{barticle}
\bauthor{\bsnm{Carlini}, \binits{A.}},
\bauthor{\bsnm{Bobin}, \binits{C.}},
\bauthor{\bsnm{Paindavoine}, \binits{M.}},
\bauthor{\bsnm{Thevenin}, \binits{M.}}:
\batitle{{A methodology for alpha particles identification in liquid
  scintillation using a cost-efficient Artificial Neural Network}}.
\bjtitle{Nuclear Instruments and Methods in Physics Research Section A:
  Accelerators, Spectrometers, Detectors and Associated Equipment}
\bvolume{1064},
\bfpage{169369}
(\byear{2024})
\doiurl{10.1016/j.nima.2024.169369}
\end{barticle}
\endbibitem

\bibitem[\protect\citeauthoryear{Kim et~al.}{2019}]{KIM201983}
\begin{barticle}
\bauthor{\bsnm{Kim}, \binits{J.}},
\bauthor{\bsnm{Park}, \binits{K.}},
\bauthor{\bsnm{Cho}, \binits{G.}}:
\batitle{Multi-radioisotope identification algorithm using an artificial neural
  network for plastic gamma spectra}.
\bjtitle{Applied Radiation and Isotopes}
\bvolume{147},
\bfpage{83}--\blpage{90}
(\byear{2019})
\doiurl{10.1016/j.apradiso.2019.01.005}
\end{barticle}
\endbibitem

\bibitem[\protect\citeauthoryear{Angloher et~al.}{2023}]{Angloher2023}
\begin{barticle}
\bauthor{\bsnm{Angloher}, \binits{G.}}, \betal:
\batitle{{Towards an automated data cleaning with deep learning in CRESST}}.
\bjtitle{The European Physical Journal Plus}
\bvolume{138}(\bissue{1}),
\bfpage{100}
(\byear{2023})
\doiurl{10.1140/epjp/s13360-023-03674-2}
\end{barticle}
\endbibitem

\bibitem[\protect\citeauthoryear{Kim et~al.}{2023}]{KIM2023110880}
\begin{barticle}
\bauthor{\bsnm{Kim}, \binits{J.}},
\bauthor{\bsnm{Jeon}, \binits{B.}},
\bauthor{\bsnm{Hwang}, \binits{J.}},
\bauthor{\bsnm{Song}, \binits{G.}},
\bauthor{\bsnm{Moon}, \binits{M.}},
\bauthor{\bsnm{Cho}, \binits{G.}}:
\batitle{Pulse height estimation and pulse shape discrimination in pile-up
  neutron and gamma ray signals from an organic scintillation detector using
  multi-task learning}.
\bjtitle{Applied Radiation and Isotopes}
\bvolume{199},
\bfpage{110880}
(\byear{2023})
\doiurl{10.1016/j.apradiso.2023.110880}
\end{barticle}
\endbibitem

\bibitem[\protect\citeauthoryear{Jeon et~al.}{2022}]{9667358}
\begin{barticle}
\bauthor{\bsnm{Jeon}, \binits{B.}},
\bauthor{\bsnm{Lim}, \binits{S.}},
\bauthor{\bsnm{Lee}, \binits{E.}},
\bauthor{\bsnm{Hwang}, \binits{Y.-S.}},
\bauthor{\bsnm{Chung}, \binits{K.-J.}},
\bauthor{\bsnm{Moon}, \binits{M.}}:
\batitle{{Deep Learning-Based Pulse Height Estimation for Separation of Pile-Up
  Pulses From NaI(Tl) Detector}}.
\bjtitle{IEEE Transactions on Nuclear Science}
\bvolume{69}(\bissue{6}),
\bfpage{1344}--\blpage{1351}
(\byear{2022})
\doiurl{10.1109/TNS.2021.3140050}
\end{barticle}
\endbibitem

\bibitem[\protect\citeauthoryear{Regadío et~al.}{2021}]{REGADIO2021165403}
\begin{barticle}
\bauthor{\bsnm{Regadío}, \binits{A.}},
\bauthor{\bsnm{Esteban}, \binits{L.}},
\bauthor{\bsnm{Sánchez-Prieto}, \binits{S.}}:
\batitle{Unfolding using deep learning and its application on pulse height
  analysis and pile-up management}.
\bjtitle{Nuclear Instruments and Methods in Physics Research Section A:
  Accelerators, Spectrometers, Detectors and Associated Equipment}
\bvolume{1005},
\bfpage{165403}
(\byear{2021})
\doiurl{10.1016/j.nima.2021.165403}
\end{barticle}
\endbibitem

\bibitem[\protect\citeauthoryear{Berg and Cherry}{2018}]{Berg_2018}
\begin{barticle}
\bauthor{\bsnm{Berg}, \binits{E.}},
\bauthor{\bsnm{Cherry}, \binits{S.R.}}:
\batitle{{Using convolutional neural networks to estimate time-of-flight from
  {PET} detector waveforms}}.
\bjtitle{Physics in Medicine {\&} Biology}
\bvolume{63}(\bissue{2}),
\bfpage{02}--\blpage{01}
(\byear{2018})
\doiurl{10.1088/1361-6560/aa9dc5}
\end{barticle}
\endbibitem

\bibitem[\protect\citeauthoryear{Onishi et~al.}{2022}]{Onishi_2022}
\begin{barticle}
\bauthor{\bsnm{Onishi}, \binits{Y.}},
\bauthor{\bsnm{Hashimoto}, \binits{F.}},
\bauthor{\bsnm{Ote}, \binits{K.}},
\bauthor{\bsnm{Ota}, \binits{R.}}:
\batitle{{Unbiased TOF estimation using leading-edge discriminator and
  convolutional neural network trained by single-source-position waveforms}}.
\bjtitle{Physics in Medicine {\&} Biology}
\bvolume{67}(\bissue{4}),
\bfpage{04}--\blpage{01}
(\byear{2022})
\doiurl{10.1088/1361-6560/ac508f}
\end{barticle}
\endbibitem

\bibitem[\protect\citeauthoryear{Wu et~al.}{2023}]{10038575}
\begin{barticle}
\bauthor{\bsnm{Wu}, \binits{Q.}}, \betal:
\batitle{{PMT Waveform Timing Analysis Using Machine Learning Method}}.
\bjtitle{IEEE Transactions on Nuclear Science}
\bvolume{70}(\bissue{6}),
\bfpage{1178}--\blpage{1182}
(\byear{2023})
\doiurl{10.1109/TNS.2023.3242650}
\end{barticle}
\endbibitem

\bibitem[\protect\citeauthoryear{Ai et~al.}{2023}]{Ai_2023}
\begin{barticle}
\bauthor{\bsnm{Ai}, \binits{P.}}, \betal:
\batitle{{Label-free timing analysis of SiPM-based modularized detectors with
  physics-constrained deep learning}}.
\bjtitle{Machine Learning: Science and Technology}
\bvolume{4}(\bissue{4}),
\bfpage{045020}
(\byear{2023})
\doiurl{10.1088/2632-2153/acfd09}
\end{barticle}
\endbibitem

\bibitem[\protect\citeauthoryear{Heshmati et~al.}{2022}]{HESHMATI2022110265}
\begin{barticle}
\bauthor{\bsnm{Heshmati}, \binits{K.}},
\bauthor{\bsnm{Ghal-Eh}, \binits{N.}},
\bauthor{\bsnm{Najafabadi}, \binits{R.I.}},
\bauthor{\bsnm{Vega-Carrillo}, \binits{H.R.}}:
\batitle{Gamma-ray energy spectrum unfolding of plastic scintillators using
  artificial neural network}.
\bjtitle{Applied Radiation and Isotopes}
\bvolume{186},
\bfpage{110265}
(\byear{2022})
\doiurl{10.1016/j.apradiso.2022.110265}
\end{barticle}
\endbibitem

\bibitem[\protect\citeauthoryear{Tajik}{2024}]{TAJIK2024111375}
\begin{barticle}
\bauthor{\bsnm{Tajik}, \binits{M.}}:
\batitle{{Unfolding of mono-energy neutron spectra using artificial neural
  network based on LMBP training algorithm}}.
\bjtitle{Applied Radiation and Isotopes}
\bvolume{210},
\bfpage{111375}
(\byear{2024})
\doiurl{10.1016/j.apradiso.2024.111375}
\end{barticle}
\endbibitem

\bibitem[\protect\citeauthoryear{Jiang et~al.}{2025}]{Jiang2025}
\begin{barticle}
\bauthor{\bsnm{Jiang}, \binits{W.}},
\bauthor{\bsnm{Huang}, \binits{G.}},
\bauthor{\bsnm{Liu}, \binits{Z.}},
\bauthor{\bsnm{Luo}, \binits{W.}},
\bauthor{\bsnm{Wen}, \binits{L.}},
\bauthor{\bsnm{Luo}, \binits{J.}}:
\batitle{{Machine-learning based photon counting for PMT waveforms and its
  application to the improvement of the energy resolution in large liquid
  scintillator detectors}}.
\bjtitle{The European Physical Journal C}
\bvolume{85}(\bissue{1}),
\bfpage{69}
(\byear{2025})
\doiurl{10.1140/epjc/s10052-024-13724-3}
\end{barticle}
\endbibitem

\bibitem[\protect\citeauthoryear{Ai et~al.}{2021}]{Ai_2021}
\begin{barticle}
\bauthor{\bsnm{Ai}, \binits{P.}},
\bauthor{\bsnm{Deng}, \binits{Z.}},
\bauthor{\bsnm{Wang}, \binits{Y.}},
\bauthor{\bsnm{Li}, \binits{L.}}:
\batitle{{Neural network-featured timing systems for radiation detectors:
  performance evaluation based on bound analysis}}.
\bjtitle{Journal of Instrumentation}
\bvolume{16}(\bissue{09}),
\bfpage{09019}
(\byear{2021})
\doiurl{10.1088/1748-0221/16/09/p09019}
\end{barticle}
\endbibitem

\bibitem[\protect\citeauthoryear{Wu et~al.}{2023}]{DBLP:conf/iclr/WuHLZ0L23}
\begin{bchapter}
\bauthor{\bsnm{Wu}, \binits{H.}},
\bauthor{\bsnm{Hu}, \binits{T.}},
\bauthor{\bsnm{Liu}, \binits{Y.}},
\bauthor{\bsnm{Zhou}, \binits{H.}},
\bauthor{\bsnm{Wang}, \binits{J.}},
\bauthor{\bsnm{Long}, \binits{M.}}:
\bctitle{{TimesNet: Temporal 2D-Variation Modeling for General Time Series
  Analysis}}.
In: \bbtitle{The Eleventh International Conference on Learning Representations,
  {ICLR} 2023, Kigali, Rwanda, May 1-5, 2023}
(\byear{2023}).
\bcomment{arXiv:2210.02186}
\end{bchapter}
\endbibitem

\bibitem[\protect\citeauthoryear{Chen and He}{2021}]{DBLP:conf/cvpr/ChenH21}
\begin{bchapter}
\bauthor{\bsnm{Chen}, \binits{X.}},
\bauthor{\bsnm{He}, \binits{K.}}:
\bctitle{{Exploring Simple Siamese Representation Learning}}.
In: \bbtitle{{IEEE} Conference on Computer Vision and Pattern Recognition,
  {CVPR} 2021, Virtual, June 19-25, 2021},
pp. \bfpage{15750}--\blpage{15758}
(\byear{2021}).
\doiurl{10.1109/CVPR46437.2021.01549}
\end{bchapter}
\endbibitem

\bibitem[\protect\citeauthoryear{Akerib et~al.}{2013}]{AKERIB2013111}
\begin{barticle}
\bauthor{\bsnm{Akerib}, \binits{D.S.}}, \betal:
\batitle{{The Large Underground Xenon (LUX) experiment}}.
\bjtitle{Nuclear Instruments and Methods in Physics Research Section A:
  Accelerators, Spectrometers, Detectors and Associated Equipment}
\bvolume{704},
\bfpage{111}--\blpage{126}
(\byear{2013})
\doiurl{10.1016/j.nima.2012.11.135}
\end{barticle}
\endbibitem

\bibitem[\protect\citeauthoryear{Akerib et~al.}{2018}]{PhysRevD.97.112002}
\begin{barticle}
\bauthor{\bsnm{Akerib}, \binits{D.S.}}, \betal:
\batitle{{Liquid xenon scintillation measurements and pulse shape
  discrimination in the LUX dark matter detector}}.
\bjtitle{Phys. Rev. D}
\bvolume{97},
\bfpage{112002}
(\byear{2018})
\doiurl{10.1103/PhysRevD.97.112002}
\end{barticle}
\endbibitem

\bibitem[\protect\citeauthoryear{{The MPD Collaboration}}{}]{NICA-MPD-ECAL-TDR}
\begin{botherref}
\oauthor{\bsnm{{The MPD Collaboration}}}:
{MPD} {NICA} Technical Design Report of the Electromagnetic calorimeter
  ({ECal}).
\url{http://mpd.jinr.ru/wp-content/uploads/2019/01/TDR\_ECAL\_v3.6\_2019.pdf}
\end{botherref}
\endbibitem

\bibitem[\protect\citeauthoryear{{The MPD Collaboration}}{}]{NICA-MPD-CDR}
\begin{botherref}
\oauthor{\bsnm{{The MPD Collaboration}}}:
{The MultiPurpose Detector -- MPD (Conceptual Design Report)}.
\url{http://mpd.jinr.ru/wp-content/uploads/2016/04/MPD\_CDR\_en.pdf}
\end{botherref}
\endbibitem

\bibitem[\protect\citeauthoryear{Pedregosa et~al.}{2011}]{scikit-learn}
\begin{barticle}
\bauthor{\bsnm{Pedregosa}, \binits{F.}}, \betal:
\batitle{{Scikit-learn: Machine Learning in {P}ython}}.
\bjtitle{Journal of Machine Learning Research}
\bvolume{12},
\bfpage{2825}--\blpage{2830}
(\byear{2011})
\end{barticle}
\endbibitem

\bibitem[\protect\citeauthoryear{Zhou et~al.}{2021}]{Zhou:21}
\begin{barticle}
\bauthor{\bsnm{Zhou}, \binits{X.}},
\bauthor{\bsnm{Bec}, \binits{J.}},
\bauthor{\bsnm{Yankelevich}, \binits{D.}},
\bauthor{\bsnm{Marcu}, \binits{L.}}:
\batitle{Multispectral fluorescence lifetime imaging device with a silicon
  avalanche photodetector}.
\bjtitle{Opt. Express}
\bvolume{29}(\bissue{13}),
\bfpage{20105}--\blpage{20120}
(\byear{2021})
\doiurl{10.1364/OE.425632}
\end{barticle}
\endbibitem

\bibitem[\protect\citeauthoryear{De~Geronimo et~al.}{2025}]{10857362}
\begin{barticle}
\bauthor{\bsnm{De~Geronimo}, \binits{G.}},
\bauthor{\bsnm{Zhu}, \binits{Y.}},
\bauthor{\bsnm{Berry}, \binits{J.E.}},
\bauthor{\bsnm{He}, \binits{Z.}}:
\batitle{{Waveform-Sampling Front-End ASIC for 3-D Position-Sensitive
  Detectors}}.
\bjtitle{IEEE Transactions on Nuclear Science}
\bvolume{72}(\bissue{3}),
\bfpage{925}--\blpage{935}
(\byear{2025})
\doiurl{10.1109/TNS.2025.3536220}
\end{barticle}
\endbibitem

\bibitem[\protect\citeauthoryear{Topko et~al.}{2022}]{TOPKO2022166680}
\begin{barticle}
\bauthor{\bsnm{Topko}, \binits{Y.}},
\bauthor{\bsnm{Topko}, \binits{B.}},
\bauthor{\bsnm{Khabarov}, \binits{S.}},
\bauthor{\bsnm{Zamyatin}, \binits{N.}}:
\batitle{{SoC-FPGA based data acquisition system for position sensitive silicon
  detectors}}.
\bjtitle{Nuclear Instruments and Methods in Physics Research Section A:
  Accelerators, Spectrometers, Detectors and Associated Equipment}
\bvolume{1033},
\bfpage{166680}
(\byear{2022})
\doiurl{10.1016/j.nima.2022.166680}
\end{barticle}
\endbibitem

\bibitem[\protect\citeauthoryear{Ai et~al.}{2022}]{Ai_2022}
\begin{barticle}
\bauthor{\bsnm{Ai}, \binits{P.}},
\bauthor{\bsnm{Deng}, \binits{Z.}},
\bauthor{\bsnm{Wang}, \binits{Y.}},
\bauthor{\bsnm{Shen}, \binits{C.}}:
\batitle{Universal uncertainty estimation for nuclear detector signals with
  neural networks and ensemble learning}.
\bjtitle{Journal of Instrumentation}
\bvolume{17}(\bissue{02}),
\bfpage{02032}
(\byear{2022})
\doiurl{10.1088/1748-0221/17/02/P02032}
\end{barticle}
\endbibitem

\bibitem[\protect\citeauthoryear{Mishkin
  et~al.}{2018}]{DBLP:conf/nips/MishkinKNSK18}
\begin{bchapter}
\bauthor{\bsnm{Mishkin}, \binits{A.}},
\bauthor{\bsnm{Kunstner}, \binits{F.}},
\bauthor{\bsnm{Nielsen}, \binits{D.}},
\bauthor{\bsnm{Schmidt}, \binits{M.}},
\bauthor{\bsnm{Khan}, \binits{M.E.}}:
\bctitle{{SLANG: Fast Structured Covariance Approximations for Bayesian Deep
  Learning with Natural Gradient}}.
In: \bbtitle{Advances in Neural Information Processing Systems 31: Annual
  Conference on Neural Information Processing Systems 2018, NeurIPS 2018,
  December 3-8, 2018, Montr{\'{e}}al, Canada},
pp. \bfpage{6248}--\blpage{6258}
(\byear{2018})
\end{bchapter}
\endbibitem

\bibitem[\protect\citeauthoryear{Ansel
  et~al.}{2024}]{DBLP:conf/asplos/AnselYHGJVBBBBC24}
\begin{bchapter}
\bauthor{\bsnm{Ansel}, \binits{J.}}, \betal:
\bctitle{{PyTorch 2: Faster Machine Learning Through Dynamic Python Bytecode
  Transformation and Graph Compilation}}.
In: \bbtitle{Proceedings of the 29th {ACM} International Conference on
  Architectural Support for Programming Languages and Operating Systems, Volume
  2, {ASPLOS} 2024, La Jolla, CA, USA, 27 April 2024- 1 May 2024},
pp. \bfpage{929}--\blpage{947}
(\byear{2024}).
\doiurl{10.1145/3620665.3640366}
\end{bchapter}
\endbibitem

\bibitem[\protect\citeauthoryear{Kingma and
  Ba}{2015}]{DBLP:journals/corr/KingmaB14}
\begin{bchapter}
\bauthor{\bsnm{Kingma}, \binits{D.P.}},
\bauthor{\bsnm{Ba}, \binits{J.}}:
\bctitle{{Adam: A Method for Stochastic Optimization}}.
In: \bbtitle{3rd International Conference on Learning Representations, {ICLR}
  2015, San Diego, CA, USA, May 7-9, 2015, Conference Track Proceedings}
(\byear{2015}).
\bcomment{arXiv:1412.6980}
\end{bchapter}
\endbibitem

\end{thebibliography}

\end{document}